\documentclass[11pt]{article}

\usepackage[utf8x]{inputenc}
\usepackage[english]{babel}
\usepackage{amsmath,amssymb}
\usepackage{graphicx}
\usepackage{multicol}
\usepackage{xspace}
\usepackage[a4paper,nohead,twoside]{geometry}
\usepackage[bookmarks=true,
            bookmarksopen=true,colorlinks=true,breaklinks=true,linkcolor=blue,%
            citecolor=blue]{hyperref}
\usepackage[font=small,format=plain,labelfont=bf,up,textfont=it,up]{caption} 
\usepackage[affil-it,max2]{authblk}

\hypersetup{%
  pdftitle = {Graceful exit from inflation for minimally coupled Bianchi A scalar field models},
  pdfsubject = {},
  pdfauthor = {F. Beyer and L. Escobar},
  pdfkeywords = {}%
}

\newcommand{\Eqref}[1]{Eq.~\eqref{#1}}
\newcommand{\Eqsref}[1]{Eqs.~\eqref{#1}}
\newcommand{\Sectionref}[1]{Section~\ref{#1}}

\newcommand{\Figref}[1]{Fig.~\ref{#1}}
\newcommand{\Figsref}[1]{Figs.~\ref{#1}}
\newcommand{\keyword}[1]{\textit{#1}\xspace}

\title{ \textbf{Graceful exit from inflation for minimally coupled Bianchi A scalar field models} }
\author[]{F. Beyer\footnote{Email: fbeyer@maths.otago.ac.nz.} }
\author[]{L. Escobar\footnote{Email: lescobar@maths.otago.ac.nz.} }
\affil[]{Department of Mathematics and Statistics, University of
  Otago,\\ P.O.\ Box 56, 9054 Dunedin, New Zealand}

\date{\today}

\graphicspath{{plots/}}

\begin{document}

\maketitle

\begin{abstract}
We consider the dynamics of Bianchi A scalar field models which undergo inflation. The main question is under which conditions does inflation come to an end and is succeeded by a decelerated epoch. This so-called ``graceful exit'' from inflation is an important ingredient in the standard model of cosmology, but is, at this stage, only understood for restricted classes of solutions. We present new results obtained by a combination of analytical and numerical techniques.
\end{abstract}

\renewcommand{\figurename}{Fig.} 

\section{Introduction}

Studies of cosmological solutions of Einstein's field equations have led to astonishing insights about the history and fate of our own universe.  
In particular, the \keyword{standard model of cosmology} \cite{Mukhanov:2005ts} has been remarkably successful, for example in describing the distribution of stars and galaxies over time, the primordial nucleosynthesis and the cosmic microwave background (CMB). The standard model is based on the \keyword{cosmological principle}. The idea is that there should be no preferred place nor direction in the universe and hence it should be \keyword{spatially homogeneous and isotropic} -- at least on average over large scales. In Einstein's theory of gravity, this idea leads to the \keyword{Friedmann-Robertson-Walker (FRW) models}. However, when FRW models are coupled to ``normal'' matter fields only and the results are compared with observations, several problems arise, for example the \keyword{flatness problem} and the \keyword{horizon problem}. The idea of \keyword{inflation}, which was introduced for the first time in \cite{Guth:1981ex},  addresses these problems. Inflation is a short phase of rapid accelerated expansion just after the big bang which is driven by a hypothetic matter field called \keyword{inflaton}. Indeed, inflation has become an important cornerstone of the standard model. An introduction to this idea from a more mathematical point of view can be found in \cite{Rendall:2006dk}. However, we also point the reader to the critical discussion in \cite{Brandenberger:2012vj}.
It is generally believed that the inflationary epoch must have stopped after approximately $75-100$ e-folds and was then succeeded by an era of decelerated expansion where ``normal'' 
baryonic matter, radiation, neutrinos and dark matter became dominant. The transition from inflation to this decelerated stage is often called \keyword{graceful exit}. If a graceful exit from inflation exists then one also speaks of \keyword{transient inflation} as opposed to \keyword{eternal inflation} which lasts forever. In this paper, we focus on such transitions from inflation to a decelerated regime and hence the graceful exit problem. Notice that for us here,  the term ``inflation'' is defined simply as an epoch of accelerated expansion (in the same way as done for example in \cite{Rendall:2006dk}). In particular, we do \textit{not} make any ``slow roll'' assumptions because such approximations are not necessary for our analysis here. Indeed there is some confusion about the slow roll approximation in the literature \cite{Liddle:1994gm} and hence we decided to not discuss this. 

The standard model of cosmology is a drastic simplification since our universe is clearly not precisely spatially homogeneous and isotropic. Normally the idea is that  within the light cone of an arbitrary observer, ``initial'' inhomogeneities and anisotropies, which are present just after the big bang, die out rapidly during inflation. Then, when inflation stops and the universe becomes decelerated, these should stay small in comparison to other measurable quantities. This kind of argument is often used as a justification why the standard model is expected to be a good description of the \textit{whole} history of the universe. However, this is far from being clear in general. On the one hand, it is not clear how the averaging process which is necessary for describing the (presumably) \emph{approximately} homogeneous and isotropic universe by \emph{exactly} homogeneous and isotropic models is defined. In fact, it is possible that nonlinear effects give rise to back-reaction phenomena which are not taken into account by the standard model, see e.g.\ \cite{Coley:2010th,Buchert:2001gh}.
On the other hand, the stability of Friedmann-Robertson-Walker solutions within the space of general solutions of Einstein's equation is not understood. Only in the case of certain eternally inflationary models, the problem of stability has been solved \cite{Ringstrom:2008kx,Ringstrom:2009zz}. It may therefore be the case that the standard model does not describe the whole history of our universe reliably. While there is relatively firm support for the idea that inhomogeneities and anisotropies indeed decay \textit{during} inflation, it is  particularly unclear what happens when inflation \textit{comes to an end}. In order to address such fundamental questions, we need in principle to give up homogeneity and isotropy completely and study the dynamics of general classes of models. In practice it is, however, very difficult to investigate completely general solutions.
In this paper here, we compromise by keeping the assumption of spatial homogeneity but give up isotropy. In this setting, the field equations imply ``only'' \textit{ordinary} differential equations (plus algebraic constraints), which are relatively tractable. 

The simplest models for accelerated expansion are obtained by introducing a positive cosmological constant to Einstein's field equations\footnote{We assume the signature $(-+++)$ for the spacetime metric. Our sign convention is that the de-Sitter spacetime is a solution of Einstein's vacuum equations with a \textit{positive} cosmological constant.} and otherwise restrict to ``normal'' matter fields, i.e., matter fields which satisfy the dominant and strong energy conditions. A theorem by Wald \cite{Wald:1983if} (which applies to the spatially homogeneous case) states that inflation is \keyword{eternal}  for such models and  there is therefore no graceful exit. Hence models with a cosmological constant cannot describe the transition from inflation to ``ordinary'' matter-dominated dynamics. In this paper, we study minimally coupled scalar field\footnote{Without further notice we assume that the scalar field is real.} models. For such models, as we will describe in more detail later, we are 
free 
to choose the \keyword{scalar field potential} $V$ as a smooth function of the scalar field $\phi$. There is a lot of freedom for this choice, but physical considerations give rise to some restrictions. 
The general idea is that, during the evolution of the universe, the scalar field ``rolls down'' the potential hill while it experiences a friction force whose magnitude is determined by the geometry of the spacetime. This suggests that the choice of the shape of the potential has important consequences for the evolution.
Later on, we shall describe the current knowledge about the dynamics of Bianchi scalar field models. 
In turns out that most of the works on scalar field models with accelerated expansion in the literature are targeted to situations where inflation does \textit{not} stop and hence there is no graceful exit. This is so because the main interest of many authors lies in another fundamental outstanding problem of mathematical cosmology: the \textit{cosmic no-hair conjecture}, which was formulated first in \cite{gibbons77,Hawking:1982ir}, see also \cite{Beyer:2009vm}. This conjecture states that generic expanding solutions of Einstein's field equations with reasonable matter fields and with a positive cosmological constant or a suitable scalar field approach the  {de-Sitter solution} asymptotically.  
Nevertheless, even though these results exclude the actual case of our interest for this paper, namely inflation with a graceful exit, they give us valuable insights about the graceful exit problem. 
While we are aware of only a few works where the emphasis lies on the graceful exit (among those are \cite{Parsons:1995bb,Barrow:2000kl,Blais:2004bj}, see also a simple example in \cite{Mukhanov:2005ts}), most of which are restricted to the homogeneous and isotropic case, we consider this problem now in the anisotropic case. We choose a specific scalar field potential here (as discussed later) which to our knowledge has not been considered in this context before.  We point out that in this paper, we are mainly interested in general qualitative properties of cosmological models as opposed to quantitative astrophysically relevant results.
Our main motivation is to systematically study a tractable, but non-trivial class of solutions of the field equations and then to ask the question, whether the concepts of inflation and graceful exit are \textit{in principle} compatible with Einstein's theory.  The purpose is therefore not so much to actually ``model'' a specific physical situation. 
 In order to make our discussion here as simple as possible and henceforth to be able to study our questions in a ``clean'' setting, we restrict to cosmological models where only one minimally coupled scalar field and no other matter fields are present. We notice that we use the word ``cosmological model'' in the sense of mathematical relativity: a cosmological model is a globally hyperbolic spacetime with compact\footnote{In the spatially homogeneous case, the topology of the Cauchy surfaces is in fact irrelevant.} Cauchy surfaces.

For our studies, the following strategy is adopted. We consider solutions of the Einstein-scalar field system from initial data in the inflationary regime. 
We show that, irrespective of the actual choice of the potential, we can always choose such initial data as long as $V(\phi_0)>0$ where $V$ is the potential and $\phi_0$ is the initial value of the scalar field.
We then study the evolution of such models by a mixture of analytical and numerical techniques in order to determine under which conditions inflation only lasts for a finite time and hence there is a graceful exit. Our results suggest that for our choice of the potential and under certain further conditions, graceful exits from inflation occur generically due to the existence of certain future attractor solutions in the decelerated regime.

The paper is organized as follows. In \Sectionref{sec:background} we discuss the necessary background for Bianchi A scalar field models and derive the full set of equations implied by Einstein's equations and the scalar field equations. In \Sectionref{sec:numericsN} we describe our numerical scheme of choosing initial data and how to deal with numerical constraint violations.
\Sectionref{sec:analysis} is the main part of the paper. In \Sectionref{sec:basicstrategy}, we summarize the main strategy. Basic consequences of our assumptions for the dynamics of the models are discussed in \Sectionref{sec:basicconsequencesN}. \Sectionref{sec:knowresults} is devoted to a summary of the known results for various classes of potentials. 
We then choose a potential for which graceful exits from inflation are, at least, not excluded by these results.
We then discuss the basic properties of this potential. It gives rise to three different cases. In one case, where the potential is strictly monotonically decreasing, we expect that the scalar field approaches infinity asymptotically and that graceful exits are possible. Hence, we analyze this case in detail in \Sectionref{sec:fixedpoints} first. If the potential is not monotonic, but has a local minimum at a finite value of $\phi$, it depends on the initial conditions whether the solutions have graceful exits. We study this case in \Sectionref{sec:nonmonotonic}.
 Finally, we close the paper with conclusions in \Sectionref{sec:discussion}.


\section{Background}
\label{sec:background}

\subsection{Self-gravitating minimally coupled scalar field models}
We consider globally hyperbolic, time-oriented oriented $4$-dimensional  smooth Lorentzian manifolds $(M,g)$ where\footnote{Our conventions for writing tensor fields are as follows. We either write the symbol of the tensor field without indices, e.g., $g$, or we use the abstract index notation, e.g., $g_{\mu\nu}$, with indices $\mu$, $\nu$,\ldots. Greek indices are therefore abstract indices and hence do in general \textit{not} refer to a coordinate basis.} $g=g_{\mu\nu}$ is a smooth Lorentzian metric.
We are concerned with solutions of Einstein's field equation  with vanishing cosmological constant (in geometrized units $c= 8 \pi G=1$ for the speed of light $c$ and the gravitational constant $G$),
\begin{equation}
\label{eq:EFE}
G_{\mu \nu}= T_{\mu\nu},
\end{equation}
where $G_{\mu\nu}$ is the Einstein tensor of $g_{\mu\nu}$ and $T_{\mu\nu}$ is the energy momentum tensor of the matter fields. As the matter field, we choose a minimally coupled scalar field $\phi$ whose energy momentum tensor is
\begin{equation}
\label{eq:energymomentumtensor}
T_{\mu\nu}= \nabla_{\mu}\phi\nabla_{\nu}\phi-g_{\mu\nu}\left(\frac{g^{\alpha\beta}}{2}\nabla_{\alpha}\phi\nabla_{\beta}\phi+V(\phi)\right).
\end{equation}
The remaining freedom is the choice of the \keyword{scalar field potential} $V(\phi)$ which is a smooth function of the scalar field $\phi$. If this potential is strictly non-negative, as we assume throughout this paper, it follows that the dominant and hence weak energy conditions are satisfied. The strong energy condition, however, may be violated. Indeed, the violation of the strong energy condition is usually considered as the essential driving force of accelerated expansion and hence inflation. The equations of motion  for the scalar field, 
\begin{equation}
\label{evolutionecuation}
\nabla^\mu\nabla_\mu\phi- \dfrac{d V(\phi)}{d \phi}=0,
\end{equation}
are derived from $\nabla_\mu{T^{\mu}}_{\nu} = 0$.
 Our particular choice of the function $V(\phi)$ is discussed later.
Notice that since no further matter fields are present, our models could also be described as \textit{self-gravitating} scalar field models.

\subsection{Bianchi A spacetimes}
In this paper, we assume a particular class of spatially homogeneous models: Bianchi models.
The basic assumption for Bianchi models is that the isometry group has a $3$-dimensional Lie subgroup which acts simply transitively by isometries on spacelike orbits $\Sigma$ in the spacetime $M$; notice that this excludes \keyword{Kantowski-Sachs models} \cite{Kantowski:1966fx,Wainwright:2005wss}. We assume that these orbits $\Sigma$ are Cauchy surfaces. Hence, $M$ is foliated by spacelike Cauchy surfaces homeomorphic to $\Sigma$ with the above symmetry property, and we let $t$ be a corresponding time function. 

Up to questions about topology, Lie groups are uniquely determined by their Lie algebras. In order to distinguish all possible Bianchi symmetries, we therefore need to classify all $3$-dimensional real Lie algebra; this is the \textit{Bianchi classification} \cite{Wainwright:2005wss}.
Let $\{\xi_1,\xi_2,\xi_3\}$ be a basis of an arbitrary $3$-dimensional real Lie algebra. The \keyword{structure constants} ${C^a}_{bc}$ with respect to this basis are defined by
\[[\xi_a,\xi_b]=\sum_{c=1}^3{C^c}_{ab} \xi_c,\quad \forall a,b=1,2,3,
\] 
where $[\cdot,\cdot]$ is the Lie bracket of the Lie algebra.
In general, there exists a symmetric matrix $(n^{ab})$ and a vector with components $(a_b)$ so that 
\begin{equation}
  \label{eq:definenanda}
{C^a}_{bc}=\sum_{e=1}^3\epsilon_{b c e}n^{ea}+a_b {\delta_c}^a-a_c {\delta_b}^a,
\end{equation} 
where $\epsilon_{abc}$ is the totally antisymmetric Levi-Civita symbol with $\epsilon_{123}=1$. 
In all of what follows, we restrict to the \keyword{Bianchi A} case, i.e., the class of \keyword{uni-modular} $3$-dimensional real Lie algebras obtained by the restriction $a_b=0$. 
Since $(n^{ab})$ is a symmetric matrix, we can assume without loss of generality a basis for which this matrix is diagonal with real eigenvalues $n_1$, $n_2$ and $n_3$. The remaining freedom to choose the basis 
gives rise to $6$ distinct classes of Bianchi A Lie algebras given in Table~\ref{tab:BianchiA}.

In order to apply the Bianchi classification to the construction of Bianchi A spacetimes, we introduce an orthonormal frame $\{e_0,e_1,e_2,e_3\}$. The vector field $e_0$ is chosen as the future directed unit normal of the symmetry hypersurfaces $\Sigma_t$ given by $t=\text{const}$  where $t$ is the time function above. The spatial frame vectors $\{e_1,e_2,e_3\}$ are therefore tangential to each $\Sigma_t$.  As
discussed in \cite{Wainwright:2005wss}, it follows that $e_0$ is a geodesic twist-free  vector field. We can pick local coordinates $(t,x^\alpha)$ on the spacetime so that $(x^\alpha)$ are local coordinates on each leaf $\Sigma_t$ and $e_0=\partial_t$. 
On each $\Sigma_t$, we attempt to choose the spatial frame vectors such that $[e_a,\zeta_b]=0$ for all $a,b=1,2,3$, where $\zeta_1$, $\zeta_2$, $\zeta_3$ is a basis of the Lie algebra of Killing vector fields associated with the Bianchi symmetry. In this case, the orthonormal frame is called  \keyword{symmetry group invariant}. Since we solve Einstein's equations as an initial value problem below, we can, a priori, only guarantee that the orthonormal frame is symmetry group invariant on the initial hypersurface $\Sigma_0$. It is then a matter of transporting the frame appropriately during the evolution to guarantee that symmetry group invariance is preserved.
If the orthonormal frame is symmetry group invariant on $\Sigma_t$, then $\{e_1,e_2,e_3\}$ span a $3$-dimensional real Lie algebra there which is isomorphic to Lie algebra spanned by $\{\zeta_1,\zeta_2,\zeta_3\}$. We assume that the Bianchi type of this algebra is one of the types in Table~\ref{tab:BianchiA}. The matrix $(n^{ab})$, which we use in the following, is the one associated with this Lie algebra.
One can show that under suitable assumptions below, the Einstein-scalar field equations imply that the Bianchi type is preserved during the evolution, i.e., if we prescribe initial data of a particular Bianchi type, then the corresponding solution has the same Bianchi type at all times  of the evolution. It makes therefore sense to speak of the \keyword{Bianchi type of the solution}.

From now on, tensor indices, $a,b,\ldots=1,2,3$, represent components of tensor fields with respect to the spatial basis vectors $\{e_1,e_2,e_3\}$. 

\begin{table}[t]
  \centering
  \begin{tabular}{|l|l|}\hline
    Bianchi I & $n_1=n_2=n_3=0$\\
    Bianchi II & $n_2=n_3=0$, $n_1>0$\\
    Bianchi VI$_0$ &$n_1=0$, $n_2>0$, $n_3<0$\\
    Bianchi VII$_0$ & $n_1=0$, $n_2>0$, $n_3>0$\\
    Bianchi VIII & $n_1<0$, $n_2>0$, $n_3>0$\\
    Bianchi IX & $n_1>0$, $n_2>0$, $n_3>0$\\\hline
  \end{tabular}
  \caption{The Bianchi A classes. The quantities  $n_1$, $n_2$ and $n_3$ are the eigenvalues of the matrix $(n^{ab})$ of the tensor defined in \Eqref{eq:definenanda}.}
  \label{tab:BianchiA}
\end{table}

\subsection{The full system of equations}
\label{sec:finalequations}

Let us now discuss the full set of equations for the minimally coupled self-gravitating Bianchi A scalar field case now. With the choice of the frame $\{e_0,e_a\}$ as above, we introduce the \keyword{Hubble scalar} $H$ and the \keyword{shear tensor}
$\sigma_{ab}$ from the identity
  $\nabla_{e_a}(e_0)_b=H h_{ab}+\sigma_{ab}$,
making use of the fact that $e_0$ is a geodesic twist-free timelike vector field. Here $h_{ab}$ is the induced spatial metric on the $t=const$-surfaces. The shear tensor $\sigma_{ab}$ is symmetric, tracefree and spatial, i.e., $\sigma_{\mu\nu} e_0^\mu=0$.
The dimension of the Hubble scalar $H$ is (time)$^{-1}$. Without going into the details now, the full system of equations can be obtained very similarly as for the perfect fluid case in \cite{Wainwright:2005wss} if, firstly, one uses the remaining gauge freedom in the same way as there, and, secondly, one assumes that $\nabla_\mu\phi$ is a timelike non-vanishing covector field. Then it indeed makes sense to consider the scalar field as a perfect fluid with energy density and pressure
\[\mu=-\frac 12\nabla_\mu\phi\nabla^\mu\phi+V(\phi),\quad p=-\frac12 \nabla_\mu\phi\nabla^\mu\phi-V(\phi),\]
and hence with ``equation of state parameter''
\[\gamma=1+\frac{-\frac12 \nabla_\mu\phi\nabla^\mu\phi-V(\phi)}{-\frac12 \nabla_\mu\phi\nabla^\mu\phi+V(\phi)},\]
defined by the relation $p=(\gamma-1)\mu$. The analogy with the perfect fluid, however, only goes as far as this; in particular, we are not allowed to \textit{choose} an equation of state freely because the quantities $\mu$ and $p$ for a scalar field have to be considered as independent. In any case, 
it is possible to derive the full system of equations for the scalar field case also directly without using the analogy with the perfect fluid. Then it is not necessary to make the a-priori assumption above that $\nabla_\mu\phi$ is timelike (but notice that this is indeed a natural assumption in the Bianchi case).

As discussed in detail in \cite{Wainwright:2005wss}, the resulting full set of quantities, which describe the Einstein-scalar field system, is $H$, $\Sigma_+$, $\Sigma_-$, $N_1$, $N_2$, $N_3$, $x$, $y$, $\phi$ and the coordinate components of the orthonormal frame vector fields; these quantities are defined as follows. Let $\sigma_1$, $\sigma_2$, $\sigma_3$ be the diagonal elements of $\sigma_{ab}$ after an appropriate choice of the orthonormal frame as discussed in \cite{Wainwright:2005wss}, and 
$\sigma_+:=\frac 12(\sigma_{2}+\sigma_{3})$,
$\sigma_-:=\frac 1{2\sqrt 3}(\sigma_{2}-\sigma_{3})$
be the ``essential'' components of this tracefree tensor field. Let $n_1$, $n_2$, $n_3$ be the diagonal elements of $n_{ab}$.
Then we define \keyword{Hubble normalized quantities} 
$\Sigma_\pm:={\sigma_\pm}/{H}$ and
$N_a:={n_a}/{H}$ \cite{Wainwright:1999it,Wainwright:2005wss}.
For the scalar field, we define 
\begin{equation}
  \label{eq:defxy}
  x:=\frac{\dot\phi}{\sqrt 6 H},\quad
  y:=\frac{\sqrt{V(\phi)}}{\sqrt 3 H},
\end{equation} 
where the latter two can be interpreted as (the square roots of) the Hubble-normalized kinetic and potential energies, respectively, of the scalar field.
These quantities are well-defined as long as $H$ never becomes zero during the evolution since we assume $V\ge 0$. If  $H>0$ initially, i.e.,  the universe expands initially, it follows that $H>0$ at least for some time of the evolution. One can show, however, see below, that for all Bianchi A models, possibly except for Bianchi IX, it follows that $H>0$ during the \textit{whole} evolution. 

The first equation implied by the Einstein-scalar field system is the \keyword{Friedmann equation} (or Hamiltonian constraint)
\begin{equation}
  \label{eq:hamiltonianconstraint}
  1=\Sigma_+^2+\Sigma_-^2+x^2+y^2+K.
\end{equation}
Here,
\begin{equation}
\label{eq:K}
K:=-\frac{{}^3R}{6H^2}=\frac
1{12}\left(N_1^2+N_2^2+N_3^2-2(N_1N_2+N_2N_3+N_3N_1)\right),
\end{equation}
where ${}^3R$ is the spatial Ricci scalar.
The Raychaudhuri equation takes the form
\begin{equation}
  \label{eq:defqqq}
  e_0(H)=\dot H=-(1+q)H^2,
\end{equation}
with the \keyword{deceleration scalar}
\begin{equation}
  \label{eq:q}
  q:=2\Sigma^2+2x^2-y^2.
\end{equation}
Here and in the following, we use the shorthand notation $\Sigma^2:=\Sigma_+^2+\Sigma_-^2$. 
Notice that the deceleration scalar $q$   measures the acceleration of the expansion of the cosmological model:  if $q>0$, then the expansion is decelerated (hence becomes slower), while if $q<0$, the expansion is accelerated (hence becomes faster).
In particular, a change of sign from negative to positive during the evolution signals a graceful exit from inflation.

In order to write down the other equations, one introduces a new time coordinate $\tau$, the so-called \keyword{Hubble time} given by
${d\tau}/{dt}=H$,
and we denote derivatives with respect to $\tau$ by a prime
$'$. Then, we find
\[\Sigma_\pm'=-(2-q)\Sigma_\pm-S_\pm,\]
where
\begin{align}
  \label{eq:Splus}
  S_+&=\frac 16\left((N_2-N_3)^2-N_1(2N_1-N_2-N_3)\right),\\
  \label{eq:Sminus}
  S_-&=\frac 1{2\sqrt 3}(N_3-N_2)(N_1-N_2-N_3).
\end{align}
Notice that $S_+$ and $S_-$ are linear combinations of the eigenvalues of the Hubble normalized tracefree part of the spatial Ricci tensor. In the same way we obtain
\begin{align*}
  N_1'&=(q-4\Sigma_+)N_1,\\
  N_2'&=(q+2\Sigma_++2\sqrt 3\Sigma_-)N_2,\\
  N_3'&=(q+2\Sigma_+-2\sqrt 3\Sigma_-)N_3.
\end{align*}

Next we extract equations from \Eqref{evolutionecuation}. Under the assumption of Bianchi symmetry, 
this equation, together with the definition of $x$ and $y$ above, gives
  \[x'=x(q-2)-F(\phi)y^2,\quad y'=F(\phi)x y+y(1+q),\]
where we define
\begin{equation}
\label{eq:defFFirst}
  F(\phi):=\sqrt{\frac 32}\,\frac{V'(\phi)}{V(\phi)}.
\end{equation}
The evolution equation for $\phi$ is therefore simply
$\phi'=\sqrt 6 x$.
Since we can rewrite \Eqref{eq:defqqq} as
\begin{equation}
  \label{eq:evolH}
  H'=-(1+q)H,
\end{equation}
and hence interpret this as an evolution equation for $H$ and since we can derive evolution equations for the coordinate components of the orthonormal frame easily from the condition that the connection is torsion-free (we shall refrain from writing these equations down now), we have therefore now succeeded in obtaining the complete system of equations for all unknown variables. In all of what follows, we shall focus on the main core of the evolution equations
\begin{align}
  \label{eq:evolutioneqSigma}
  \Sigma_\pm'&=-(2-q)\Sigma_\pm-S_\pm,\\
  \label{eq:evolutioneqN1}
  N_1'&=(q-4\Sigma_+)N_1,\\
  \label{eq:evolutioneqN2}
  N_2'&=(q+2\Sigma_++2\sqrt 3\Sigma_-)N_2,\\
  \label{eq:evolutioneqN3}
  N_3'&=(q+2\Sigma_+-2\sqrt 3\Sigma_-)N_3,\\
  \label{eq:evolutioneqx}
  x'&=x(q-2)-F(\phi) y^2,\\
  \label{eq:evolutioneqy}
  y'&=F(\phi) x y+y(1+q),\\
   \label{eq:evolutioneqphi}
  \phi'&=\sqrt 6 x. 
\end{align}
This set of evolution equations is a dynamical system whose state space is spanned by $(\Sigma_+, \Sigma_-,
N_1, N_2, N_3, x, y, \phi)\in\mathbb{R}^8$. It is subject to the Friedmann constraint \Eqref{eq:hamiltonianconstraint}.
The quantity 
\begin{equation}
\label{eq:hamiltoniansurface}
G:=1- \Sigma^{2}_{+}-\Sigma^{2}_{-}-x^{2}-y^{2}-K,
\end{equation}
is a measure of the violation of this constraint. Suppose we prescribe data for the initial value problem of the evolution equations above which possibly violate the constraint. Hence $G$ does not necessarily vanish initially. For the corresponding solution of the evolution equations it is straightforward to derive an evolution equation for $G$:
\begin{equation}
\label{eq:subsidiarysystem}
G'= 2 q G,
\end{equation}
the \keyword{subsidiary system}. Among other things discussed below, this implies that if we prescribe initial data which satisfy the constraint, i.e., $G=0$ initially, and then determine the corresponding solution of the evolution equations, then the constraint is satisfied identically for all times. 

After the main set of evolution equations \Eqsref{eq:evolutioneqSigma} -- \eqref{eq:evolutioneqphi} above has been solved subject to the Friedmann constraint, \Eqref{eq:evolH} can be used to determine $H$ as a function of $\tau$. Moreover, the coordinate components of the orthonormal frame can be computed as a function of $\tau$ as described above.

In our application, the scalar field $\phi$ often approaches infinity rapidly. It is then useful to replace the variable $\phi$ by $\psi:=1/\phi$. With this we obtain the following alternative evolution equations 
\begin{align}
  \label{eq:evolutionxpsi}
  x'&=x(q-2)-F(\psi) y^2,\\
  \label{eq:evolutionypsi}
  y'&=F(\psi) x y+y(1+q),\\
  \label{eq:evolutionpsipsi}
  \psi'&=-\sqrt 6 x \psi^2.
\end{align}
Here, we use a slightly sloppy notation where the function $F$, which was defined as a function of $\phi$ above, is now considered as a function of $\psi=1/\phi$, i.e.,
\begin{equation}
  \label{eq:defF}
  F(\psi):=\sqrt{\frac 32} \frac{V'(1/\psi)}{V(1/\psi)}.
\end{equation}

Finally, we notice that the analogy with the perfect fluid mentioned before allows us to express
$\mu=3H^2(x^2-y^2)$, $p=3H^2(x^2-y^2)$. Hence the Hubble-normalized energy density $\Omega:=\mu/(3H^2)$ and the ``equation of state parameter'' $\gamma$
can be written as
\begin{equation}
\label{eq:perfectfluidquantitites}
\Omega=x^2+y^2,\quad \gamma=\frac{2x^2}{x^2+y^2}.
\end{equation}


\section{Numerical implementation}
\label{sec:numericsN}

\subsection{Initial data}

We are interested in initial data that satisfy three conditions: (i) They must satisfy the constraint \Eqref{eq:hamiltonianconstraint} and, (ii) they should represent an inflationary epoch, i.e., we should have $H>0$ (expansion) and the quantity $q$ in \Eqref{eq:q} should be negative (inflation) initially. Finally, (iii), the initial data should be of a particular Bianchi type which is reflected in the choice of the quantities $N_a$. 

We proceed as follows. \Eqsref{eq:hamiltonianconstraint} and \eqref{eq:q} can be solved for $x^2$ and $y^2$ and we get
\[x^2=(1-3\Sigma^2-K+q)/3,\quad y^2=(2-2K-q)/3.\]
This implies the restrictions
\[1-3\Sigma^2-K+q\ge 0,\quad 2-2K-q> 0,\]
since we shall always demand that $y>0$ (this is always possible, see \Eqref{eq:defxy}; see \Sectionref{sec:knowresults} for our motivation to assume $V>0$)
and hence
\begin{equation*}
  3\Sigma^2+K-1\le q< 2(1-K).
\end{equation*}
This inequality can be satisfied for $q<0$
if and only if
\[3\Sigma^2+K<1.\]
Notice here that $2(1-K)> 0$ follows from the Friedmann constraint.
Based on these insights, we adopt the following strategy to choose inflationary initial data for our numerical evolutions:
\begin{enumerate}
\item  Choose arbitrary numbers for the initial data of $N_1$, $N_2$ and $N_3$ (which determines the Bianchi class) and for $\Sigma_{+}$ and $\Sigma_{-}$ so that 
\[3\Sigma^2+K<1.\]
\item Choose an arbitrary number $q$ in the interval
\[3\Sigma^2+K-1\le q<0.\]
\item Choose the initial data for $x$ and $y$ as
\[x=\pm \sqrt{(1-3\Sigma^2-K+q)/3},\quad y=\sqrt{(2-2K-q)/3}.\]
\item Choose an arbitrary (positive) number for the initial data of the scalar field 
$\phi$.
\end{enumerate}

In summary, this means that for any minimally coupled Bianchi A scalar field model, irrespective of the actual choice of the potential, we can always construct initial data in the inflationary regime given by some $q<0$. In fact, the only restriction is that $V(\phi)>0$ for the initial value of $\phi$ (hence $V$ does in fact not need to be strictly positive \textit{everywhere} for this).

\subsection{Numerical implementation and constraint damping}
\label{sec:constraintdampingN}

In order to solve the ordinary differential evolution equations numerically, we have used the 4th-order Runge Kutta method (non-adaptive) and the Dormand-Prince method (adaptive time stepping).

\begin{figure}[t]
  \begin{minipage}{0.49\linewidth}
    \centering
    \includegraphics[width=\textwidth]{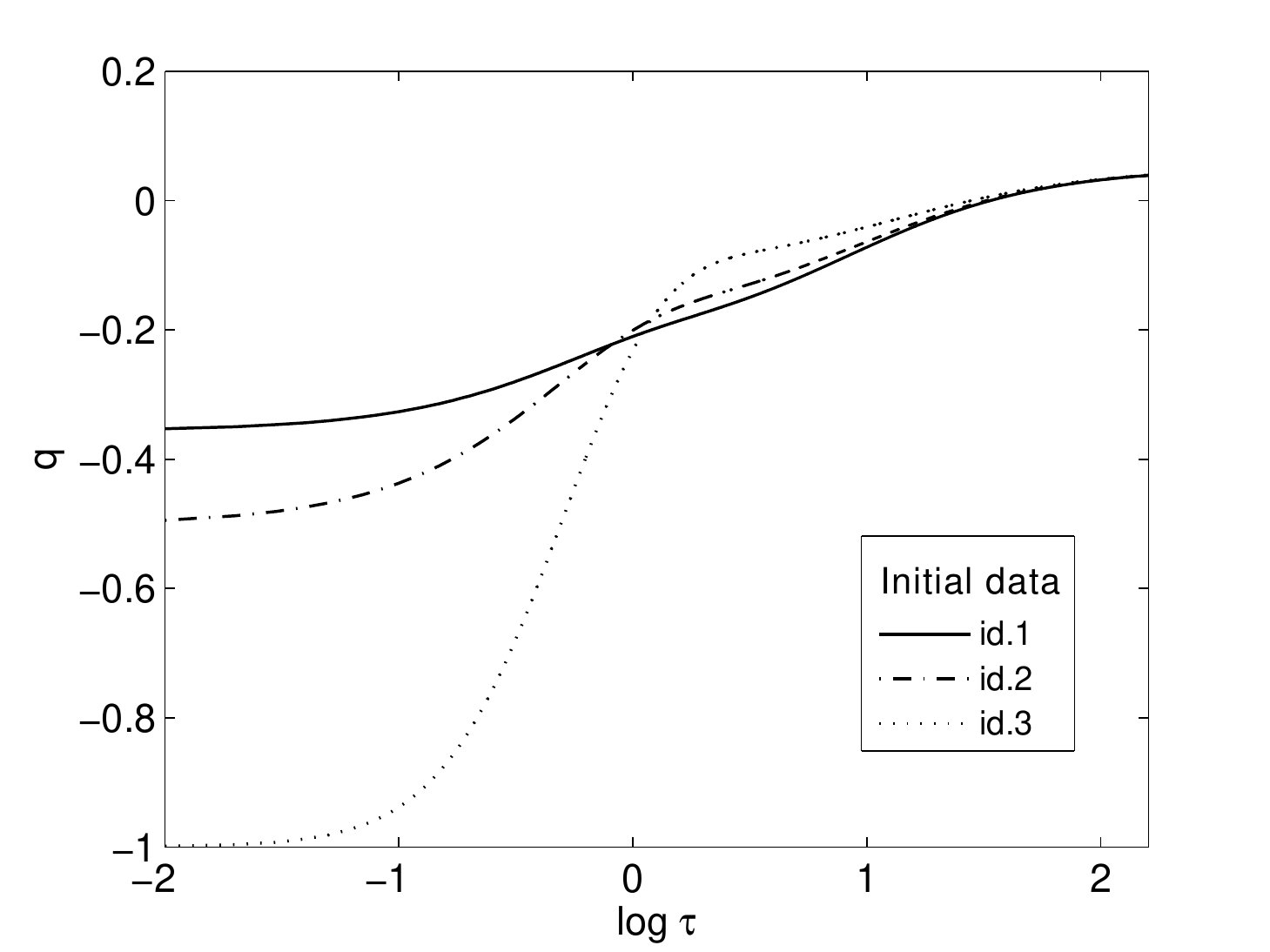}
    \caption{Deceleration scalar $q$ (BI). \vspace{0ex}}
    \label{fig:qBInotdamped}
    \
  \end{minipage}
  \begin{minipage}{0.49\linewidth}
    \centering
    \includegraphics[width=\textwidth]{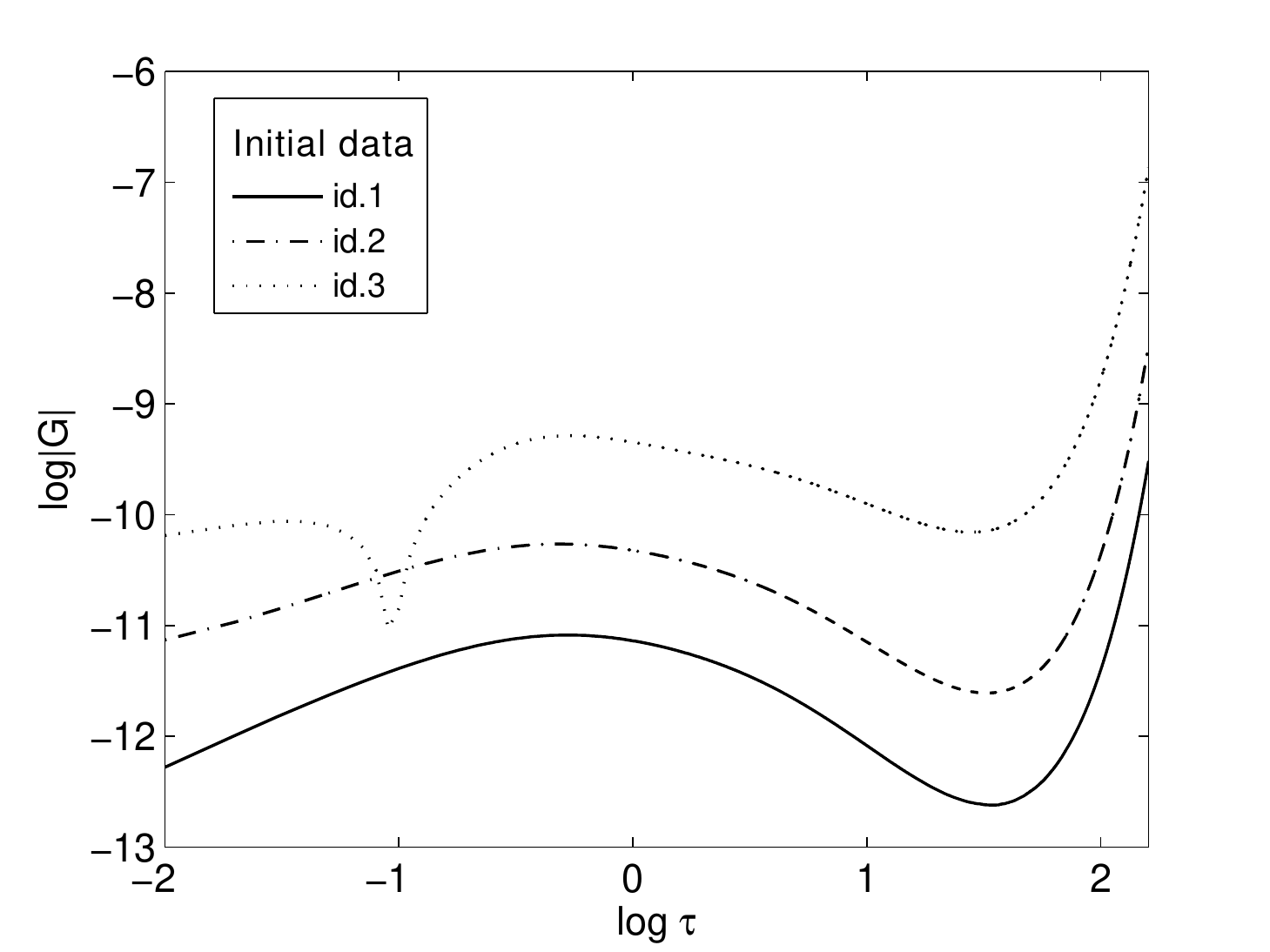}
    \caption{Constraint violations $G$ using the original evolution equations (BI).}
    \label{fig:constraintBInotdamped}
  \end{minipage}
\end{figure}

We made several numerical experiments with this setup and found quickly that the evolution equations in the form above have a serious numerical problem.
In \Figsref{fig:qBInotdamped}  and \ref{fig:constraintBInotdamped} we see that, while the constraint violation quantity $G$ (defined in \Eqref{eq:hamiltoniansurface}) is numerically well-behaved as long as $q<0$, it grows rapidly during epochs with $q>0$, i.e., when the expansion is decelerated. Eventually the numerical solutions are therefore rendered useless after short evolution times when $G$ becomes of order unity. This observation is in agreement with the subsidiary system \Eqref{eq:subsidiarysystem} of the evolution equations.
If $G$ is not exactly zero initially, we can conclude that it decays exponentially while $q<0$ (when the expansion is accelerated) and increases exponentially while $q>0$ (during decelerated epochs). 
At the beginning of the numerical calculations, $q$ is negative and hence, the constraint violations, which are non-zero initially, do not grow; the fact that they do not decay, as it is suggested by  \Eqref{eq:subsidiarysystem},  is caused by numerical discretization errors since the evolution equations are solved numerically and hence are satisfied only approximately. 

Before we present our solution to this problem, let us explain why the violation of the constraint at the initial time in \Figref{fig:constraintBInotdamped} is $\sim 10^{-11}$ , and not, as it may be expected, at the level of machine precision $\sim 10^{-15}$. The reason is simply that in all of our figures, we do not plot the initial data themselves (which correspond to $\tau=0$), but instead the solution from the first numerical evolution step onwards.

\begin{figure}[t]
    \centering
    \includegraphics[width=0.49\textwidth]{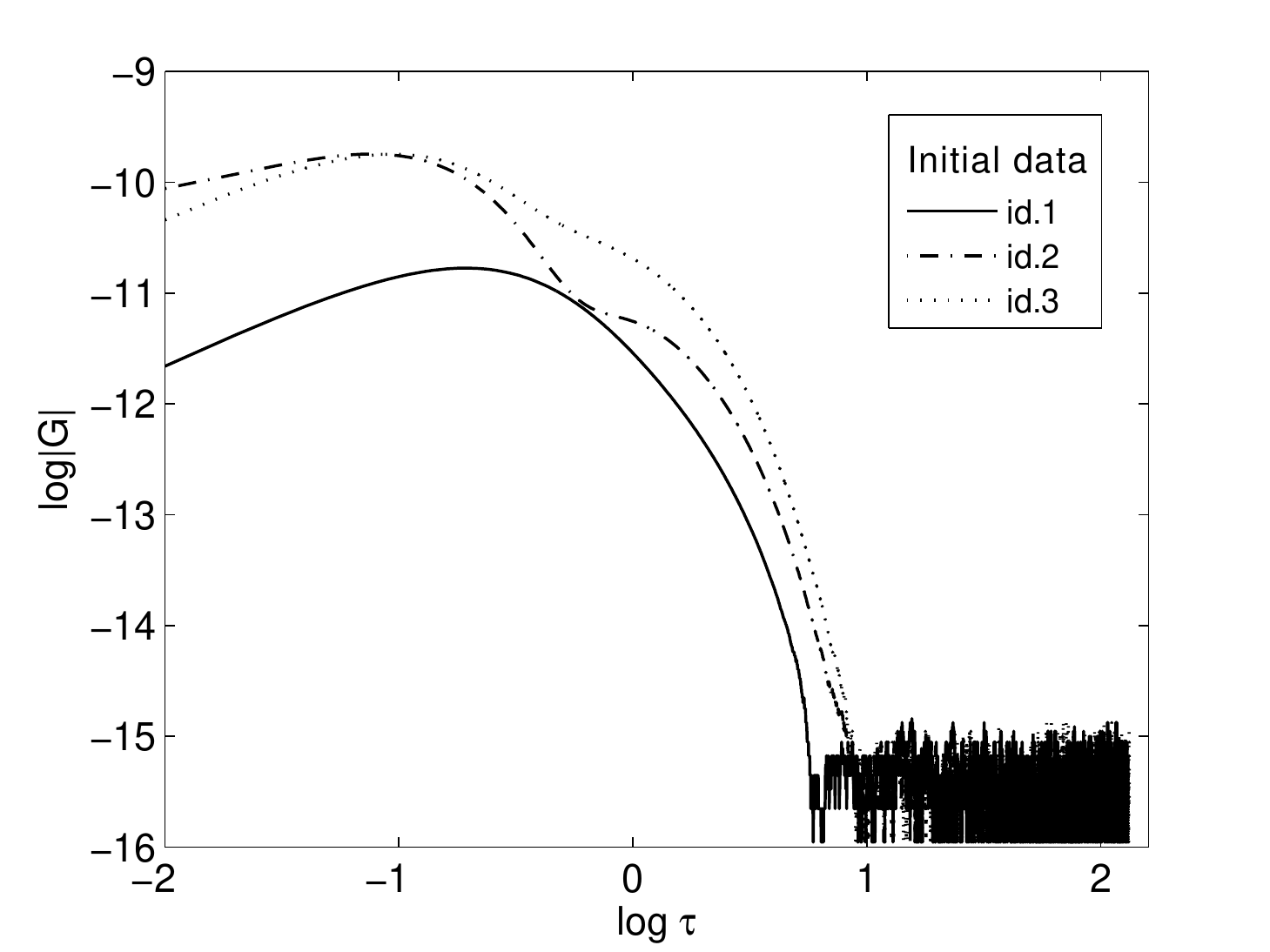}
    \caption{Constraint violations $G$ using the modified evolution equations (BI).}
    \label{const}
\end{figure}

In order to solve the instability of the constraint evolution of our dynamical system and hence to be able to make reliable long-term numerical computations, we propose to solve a system of \textit{modified} evolution equations which agrees with the original one precisely if the constraint violations vanish and which is obtained by adding suitable \keyword{constraint damping terms} to the evolution equations. The idea of damping constraint violations by modifying the evolution equations appropriately has been introduced in \cite{Brodbeck:1999en} and was then further developed for example in \cite{Gundlach05}. These modified evolution equations are 
\begin{eqnarray*}
\Sigma_{+}'&=&-(2-q)\Sigma_{\pm}-S_{+}+f_1 G,  \\
\Sigma_{-}'&=&-(2-q)\Sigma_{\pm}-S_{-}+f_2 G, \\
N_1'&=&(q-4\Sigma_+)N_1+f_3, \\
N_2'&=&(q+2\Sigma_++2\sqrt{3}\Sigma_-)N_2+f_4 G, \\
N_3'&=&(q+2\Sigma_+-2\sqrt{3}\Sigma_-)N_3+f_5 G,  \\
x'&=&x(q-2)- F(\psi) y^{2}+ f_6 G, \\
y'&=& F(\psi) x y+y(1+q)+f_7 G, \\
\psi'&=&- \sqrt{6}x \psi^2.
\end{eqnarray*}
Comparing this to the original ``unmodified'' evolution equations \Eqsref{eq:evolutioneqSigma} -- \eqref{eq:evolutioneqphi} (or more precisely  \eqref{eq:evolutionxpsi} -- \eqref{eq:evolutionpsipsi}), it becomes obvious that this modified system agrees with the original one precisely if $G=0$. Hence any solution of the original system, which satisfies the constraints, is also a solution of this modified system and vice versa. The functions $f_1,\ldots,f_7$ are so far unspecified.  The subsidiary system for these modified evolution equations is
\begin{equation*}
G'=-2\Sigma_{+} \Sigma_{+}'-2\Sigma_{-} \Sigma_{-}' -2 x x' -2 y y' - K',
\end{equation*}
which, using the evolution equations above, motivates us to choose
\begin{align*}
&f_1=2 \Sigma_{+},& &f_2=2 \Sigma_{-},& &f_3= -(N_2+N_3)/2,& &f_4=-(N_1+N_3)/2,& \\
&f_5=-(N_1+N_2)/2,& &f_6=2x,& &f_7=(\kappa-1)y,&
\end{align*}
for some constant $\kappa \geq 0$. Certainly, this choice for $f_1$,\ldots,$f_7$ is not unique, but it has the nice consequence that the subsidiary system becomes
\begin{equation}
\label{eq:modifiedsubiduarysyst}
G'=-( (N_1^2+N_2^2+N_3^2)/6 +2 \kappa  y^2) G.
\end{equation}
In particular, the factor in front of $G$ on the right hand side is negative semi-definite and hence, the constraint violations are expected to be numerically stable irrespective of the sign of $q$. In all of what follows we choose $\kappa=1$. This theoretical prediction is confirmed numerically in \Figref{const} which shows numerical evolutions for the modified system. We see that the constraint violations are now stable during the \textit{whole} evolution and are driven towards numerical double precision round off errors ($\sim 10^{-15}$). We can now hope that we are able to do  reliable long-term numerical simulations. 

The above numerical examples are Bianchi I solutions for a certain scalar field potential introduced later. We find that this way of damping constraint violations works well in all Bianchi cases as long as the numerical resolution is sufficient. Indeed, since some of the unknowns are unbounded in several Bianchi cases (more details on this later), keeping a sufficient numerical resolution can become a challenge in practice  (cf.\ e.g.\ \Figref{fig:BVIII:G}).  

A possible alternative to introducing constraint damping terms to the ``free'' evolution equations is to implement a ``constrained evolution'' scheme. For this, instead of determining the solution for \textit{all} variables from the evolution equations, one would single out one variable and instead determine it algebraically from the Friedmann constraint. By this, one would guarantee that the constraint violations are zero. In order to demonstrate, however, that the resulting numerical solution satisfies \textit{all} of Einstein's equations, we would then need to check if the one evolution equation, which has been eliminated in the first step, is satisfied. We hence arrive at the same problem as in our implementation, namely, that there is a-priori no guaranty that this additional equation is satisfied well by the numerical solution. The origin of this problem is that Einstein's equations are overdetermined and this is the same for both ``free evolution'' (as in our case) and such ``constrained evolution'' schemes.


\section{Analysis of the graceful exit problem}
\label{sec:analysis}

\subsection{Basic strategy}
\label{sec:basicstrategy}
We can restrict (without loss of generality) to expanding cosmological models characterized by $H>0$ where $H$ is the Hubble scalar. The acceleration of the cosmological model is described by the deceleration parameter $q$ in \Eqref{eq:q} which is, by convention, positive when the expansion is decelerated and negative when it is accelerated. If the solution, which corresponds to a choice of inflationary initial data as described before, has the property that $q<0$ for all times, then the inflation era is eternal. If, however, $q$ changes its sign after a finite time and stays positive, then there is a graceful exit from inflation. 
The main idea in this paper is  to identify certain future attractor solutions in the decelerated regime -- in many cases these are equilibrium points of our dynamical system.  If future attractors in the decelerated regime exist then our solutions must have the property that the initial inflationary epoch is finite and is succeeded by a decelerated regime after a graceful exit. Notice that we use the term ``attractor'' sometimes in a slightly loose sense in this paper. In the strict mathematical sense, an attractor is, in particular, a subset of the state space which is approached by generic orbits (the precise definition can be found in \cite{Wainwright:2005wss}). In the Bianchi cases VII$_0$ and VIII below, however, we refer to a particular generic asymptotic behavior as an attractor even though this asymptotic behavior cannot be associated directly with a subset of the state space.
Equilibrium points for homogeneous and isotropic scalar field models were studied before, see for instance \cite{Fang:2009hv,Kiselev:2008fw,Copeland:1998dx,Copeland:2009bi}.

\subsection{Basic consequences for the evolution}
\label{sec:basicconsequencesN}

Here we summarize some basic results about the dynamics of our models. Let us assume from now on that the potential $V$ is a strictly positive smooth function of $\phi$; this is the \textit{only} restriction on $V$ which we make in the following subsection, which will be motivated in \Sectionref{sec:knowresults}. 

Consider the Friedmann equation \Eqref{eq:hamiltonianconstraint}, which can be rewritten as
\[H^2=\sigma_+^2+\sigma_-^2+(e_0(\phi))^2/6+V(\phi)/3-{}^3 R/6,\]
after multiplication with $H^2$.
Let us assume that $H>0$ at the initial time. Hence $H$ can become non-positive during the evolution only if the right side of this equation is allowed to become zero. Under our assumptions, however, this is only possible if ${}^3 R$ can be positive. It is a standard result, see for instance \cite{Wainwright:2005wss}, which can be checked directly for instance from \Eqref{eq:K}, that ${}^3 R$ is non-positive for all Bianchi A models possibly except for Bianchi IX. For all other Bianchi A cases, we have in fact that $0\le K\le 1$. By excluding Bianchi IX for the rest of this paper, we  therefore make sure that $H$ is positive during the whole evolution and that hence our cosmological models do not recollapse. In particular, this means that the Hubble normalized variables, which we have introduced in the previous section, are well-defined during the whole evolution.

We can also show that our cosmological models must isotropize and spatial curvature must decay during inflation (i.e., while $q<0$). 
To this end, define a new quantity 
\begin{equation}
  \label{eq:defR}
  S:=1-x^2-y^2.
\end{equation}
The Friedmann constraint implies that $S=\Sigma^2+K$ and $0\le S\le 1$ (if we exclude the Bianchi IX case). The evolution equations \eqref{eq:evolutioneqSigma} -- \eqref{eq:evolutioneqphi} yield an evolution equation for $S$
\begin{equation}
  \label{eq:evolR}
  S'=2 q S-4\Sigma^2\le 2 q S.
\end{equation}
Thus, $S$, and therefore  $\Sigma^2$ and $K$, decrease rapidly during inflation ($q<0$). With somewhat more work not discussed here, one can show that in fact the full spatial curvature tensor decays during inflation. If inflation lasts forever, then this is a realization of the cosmic no-hair conjecture. We are here, however, more interested in situations when inflation does not last forever and there is a graceful exit. If inflation is sufficiently long, then our argument here suggests that anisotropies and the size of the spatial curvature are extremely small by the time of the graceful exit. One of the  questions which is of interest for us here is then: What happens when inflation is over? Do anisotropies and spatial curvature stay small or do they grow again?

Another important immediate consequence from the fact that $K\ge 0$ for all Bianchi A cases (possibly except for Bianchi IX) is that the constraint \Eqref{eq:hamiltonianconstraint} implies the boundedness of the variables $\Sigma_\pm$, $x$ and $y$. From the definition of $K$ in \Eqref{eq:K} and the conventions for the Bianchi cases in Table~\ref{tab:BianchiA}, we find that the remaining quantities have the following properties. For Bianchi I, the variables $N_a$ are identically zero and hence trivially bounded. For Bianchi II, the non-zero quantity $N_1$ must be bounded. For Bianchi VI$_0$, the two non-zero quantities $N_2$ and $N_3$ (with opposite signs) are bounded. For Bianchi VII$_0$, however, the two non-zero quantities $N_2$ and $N_3$ (with equal signs) are not necessarily bounded. For Bianchi VIII, it follows that $N_1$ must be bounded, but $N_2$ and $N_3$ can be unbounded.

We also mention here the following basic inequalities and extreme cases for the deceleration scalar $q$. The formula \Eqref{eq:q} for $q$ can be rewritten as
\begin{equation}
  \label{eq:q2}
  q=2-3y^2-2K,
\end{equation}
using the Friedmann constraint.
Hence we have $q\le 2$ for all Bianchi A cases (possibly except for Bianchi IX) and $q=2$ if and only if $K=0$ and $y=0$. We can therefore typically assume that $q<2$. Since $y^2+K\le 1$ (from the Friedmann constraint), it also follows that $q\ge -1$. The case $q=-1$ corresponds to de-Sitter like exponential inflation.

\subsection{Known results and our choice of the scalar field potential}
\label{sec:knowresults}

Let us now list the main relevant rigorous results for minimally coupled Bianchi A scalar field models for various classes of potentials. We shall use those insights in order to choose a concrete class of potentials for our numerical investigations for which solutions with graceful exits from inflation are possible.

Recall that the assumption that the potential is non-negative guarantees that  the dominant and hence weak energy conditions are satisfied. Let us therefore restrict to potentials which are non-negative. The first relevant result has been obtained by Rendall \cite{Rendall:2004gu}. If the potential has a positive lower bound (and certain further technical conditions are satisfied), then the solutions undergo eternal exponential de-Sitter like expansion asymptotically. Hence potentials with a positive lower bound do not allow solutions with a graceful exit from inflation. Since we want to avoid the complication of chaotic inflation associated with potentials with a \textit{zero local minimum} \cite{Moss:1986el,Amin:2012ho}, we must therefore focus on strictly positive potentials which approach zero in the limit $\phi\rightarrow\infty$. This could be an exponential potential \cite{Halliwell:1987de,Kitada:1993wm,Barreiro:2000ij}, in which case it was shown that eternal power-law expansion occurs if the exponent is not too negative (we shall return to this potential later but it will not be our main choice). Another possibility is an ``intermediate'' potential. The investigations in \cite{Rendall:817462} (see in particular Theorem~4 there) yield that one gets eternal ``intermediate inflation'' if, in particular, $\lim_{\phi\rightarrow\infty}V'(\phi)/V(\phi)\rightarrow 0$. See also \cite{Barrow:2007fv}. We must therefore consider potentials with the property
$\lim_{\phi\rightarrow\infty}V'(\phi)/V(\phi)< 0$ in order to make solutions with graceful exits from inflation possible.

Now the concrete model, which we have decided to study in this paper, is
the potential introduced by Albrecht-Skordis \cite{Albrecht:2000fx} in the context of FRW cosmological models with dark energy in string theory
\begin{equation}
\label{eq:potential}
V(\phi)= e^{-c_1 \phi} (c_2 + \phi^2 ),
\end{equation}
where $c_1>0$, $c_2\ge 0$ are constants.  In particular, the function $F(\psi)$ (see \Eqref{eq:defF} where $\phi$ has been replaced by $\phi=1/\psi$) has the form
\begin{equation}
  \label{eq:FpsiPotential}
  F(\psi) =\sqrt{\dfrac{3}{2}} \left( -c_1 + \dfrac{ 2 \psi }{1+c_2\psi^{2} }\right).
\end{equation}
Its limit
\begin{equation}
\label{eq:defFlimit}
 \lim_{\psi\rightarrow 0}F(\psi) = - \sqrt{\dfrac{3}{2}} c_1=:F_0<0,
\end{equation}
and hence one of the sufficient conditions for \textit{eternal} intermediate inflation in  \cite{Rendall:817462} mentioned above is in general violated.

\begin{figure}[t]
  \centering
  \includegraphics[width=0.49\textwidth]{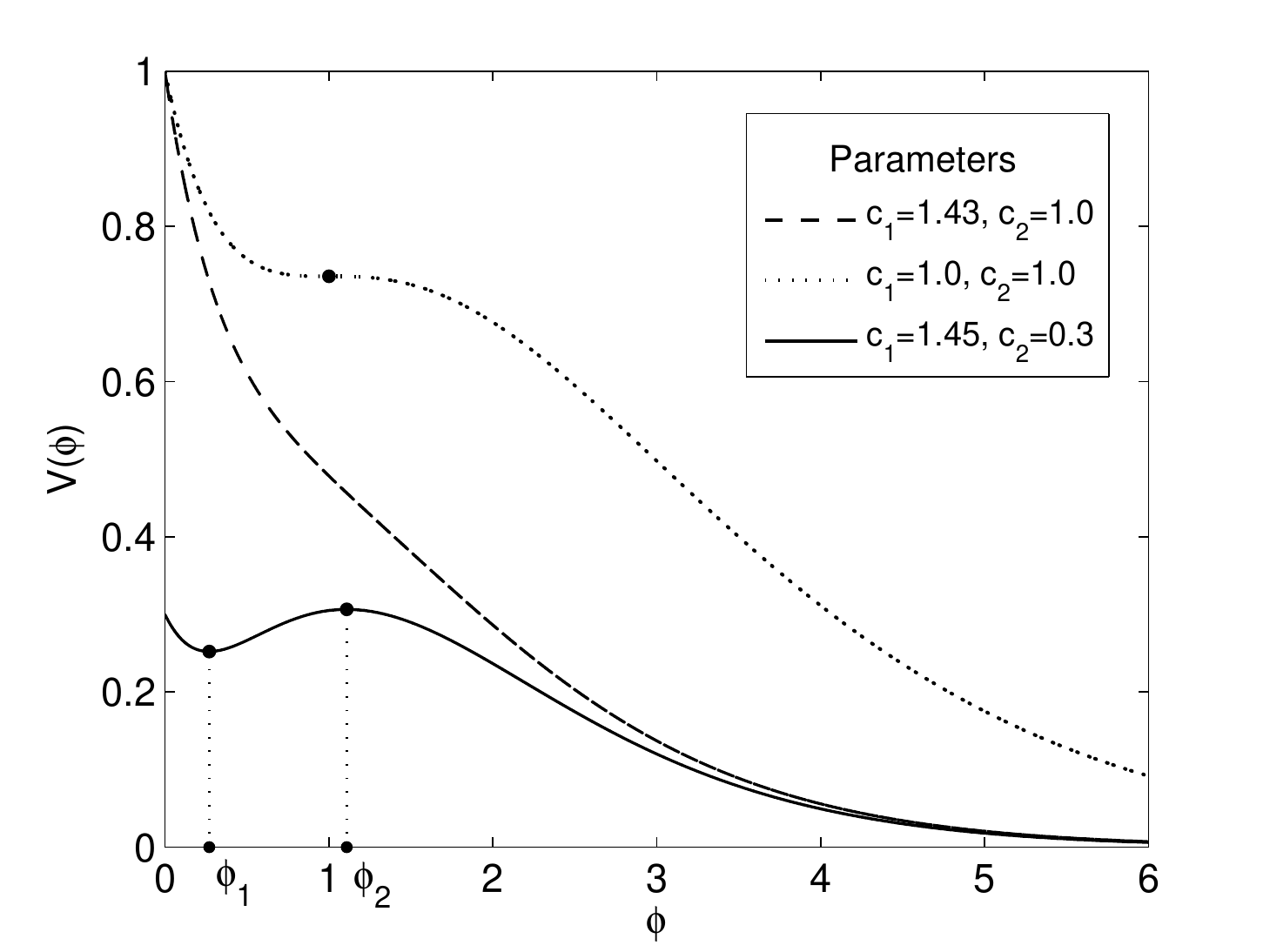}
  \caption{The three main cases for the scalar field potential $V(\phi)$. }
  \label{fig:potential}
\end{figure}
As shown in \Figref{fig:potential}, the potential \Eqref{eq:potential} gives rise to three different cases. According to the idea that the scalar field essentially ``roles down the potential hill'' during the evolution, we expect that the scalar field either goes to infinity asymptotically (in which case the discussion above suggests that a graceful exit from inflation could be possible), or is stuck  in the local minimum of the potential where we then expect eternal de-Sitter like inflation in agreement with \cite{Rendall:2004gu}. Since we are  interested in the graceful exit problem, we shall start our investigations for cases when $c_1$ and $c_2$ are such that the potential is strictly monotonic. One shows easily that this is the case when $c^{2}_{1} c_2>1$. Later in \Sectionref{sec:nonmonotonic}, we present  numerical investigations for other choices of the parameters of the potential, in particular when the potential has a ``valley''.

\subsection{The graceful exit problem for the monotonic case of the potential}
\label{sec:fixedpoints}

\subsubsection{Bianchi I \texorpdfstring{($N_1=N_2=N_3=0$)}{(N1=N2=N3=0)}}
Let us start our discussion with the simplest case: Bianchi I. This is characterized by the condition that $N_1$, $N_2$ and $N_3$ vanish identically which implies spatial flatness.

\paragraph{Future attractors.}
We first identify a Bianchi I equilibrium point of the dynamical system, i.e., a solution of the evolution equations \Eqsref{eq:evolutioneqSigma} -- \eqref{eq:evolutioneqN3}, \eqref{eq:evolutionxpsi} -- \eqref{eq:evolutionpsipsi} and of the constraint \Eqref{eq:hamiltonianconstraint} with the property 
\[\Sigma_\pm'=N_1'=N_2'=N_3'=x'=y'=\psi'=0.\] 
We determine this equilibrium point under the condition $\psi=0$ (which represent the limit $\phi\rightarrow\infty$ when the scalar field has ``completely rolled down the potential hill'' asymptotically) and for $y>0$. Notice that $y>0$ is suggested by the numerical solutions below (but not $y=0$). 
In the Bianchi I case, the condition $\Sigma_\pm'=0$ yields $\Sigma_\pm=0$ (see \Eqref{eq:evolutioneqSigma}) since $S_\pm=0$ and since $q<2$ for $y>0$ (see \Sectionref{sec:basicconsequencesN}). We can use the Friedmann constraint to express $y$ in terms of $x$, i.e., $y=\sqrt{1-x^2}$.
The evolution equation for $x$ with $x'=0$ can now be solved for $x$. This algebraic equation yields two solutions of which only one is relevant. The evolution equation for $y$ with $y'=0$ and $y=\sqrt{1-x^2}$ is then satisfied identically. We get
\begin{equation}
  \label{eq:BianchiIFP}
  \Sigma_\pm=0,\, N_1=N_2=N_3=0, \,x=-\frac{F_0}3,\, y=\sqrt{1-\frac{F_0^2}9},\, \psi=0,
\end{equation}
with $F_0$ given by \Eqref{eq:defFlimit}, i.e.,
\[F_0=- \sqrt{\dfrac{3}{2}} c_1.\]
This is the equilibrium point, which is called $F$ in \cite{Wainwright:2005wss} and which can be interpreted as a flat FRLW solution with a perfect fluid source given by $\Omega=1$ and equation of state parameter $\gamma=c_1^2/3$ (see \Eqref{eq:perfectfluidquantitites}).
Hence, for every fixed $c_1<\sqrt{6}$, we have a equilibrium point of type Bianchi I. Notice here that this is strictly speaking \textit{not} an actual solution of the field equations because $\psi=0$ (corresponding to $\phi=\infty$) lies outside the allowed range. This equilibrium point, however, can be interpreted meaningfully as the limit $\phi\rightarrow\infty$ of an actual solution. We shall always understand our equilibrium points in this way.
\Eqref{eq:q} yields the corresponding value of $q$:
\[q=\frac{c_1^2-2}2.\] 
It follows that $q>0$ (i.e., the equilibrium point is in the decelerated regime) if and only if
\begin{equation}
\label{eq:Bianchirestriction}
\sqrt{2}<c_1< \sqrt{6}.
\end{equation}
It is interesting to notice that this yields the inequality $2/3<\gamma<2$ which corresponds to one of the restrictions for the analysis of Bianchi A models with perfect fluid sources in \cite{Wainwright:2005wss}.

We claim now that generic Bianchi I solutions with our choice of the potential and for $c_1$ compatible with \Eqref{eq:Bianchirestriction} approach this equilibrium point in the future asymptotically and hence that it is a future attractor.
If this is the case the solution is decelerated and becomes isotropic and spatially flat. In particular, for evolutions starting from inflationary initial data, there is a graceful exit after a finite evolution time and the process of isotropization and approach towards spatial flatness, which occurs during inflation according to our general discussion in \Sectionref{sec:basicconsequencesN}, continues after inflation.

If this equilibrium point is really supposed to be a future attractor for Bianchi I models, then it must at least be future non-linearly stable. According to the standard theory of dynamical systems, we linearize the evolution equations for $\Sigma_\pm$, $x$, $y$ and $\psi$ with $N_1$, $N_2$, $N_3$ identically zero around the equilibrium point. We find that the corresponding evolution matrix has the eigenvalues\footnote{Notice that the eigenvalues and hence stability properties differ from the perfect fluid case discussed in \cite{Wainwright:2005wss} despite the close analogy with this case.} $(c_1^2-6)/2$ (three times repeated), $c_1^2-2$ (single) and $0$ (single). The equilibrium point therefore has a $3$-dimensional future stable subspace, a $1$-dimensional future unstable subspace and a $1$-dimensional center subspace under the restriction \Eqref{eq:Bianchirestriction}. We check that the $1$-dimensional future unstable subspace is transverse to the constraint hypersurface and hence is irrelevant for solutions which satisfy the constraint. 

In order to understand the meaning of the center subspace, we apply the center manifold theory; see for example the summary and references in \cite{Rendall:2008vq}.  For this discussion, we eliminate the evolution equation for $y$ and replace all occurrences of $y$ in the other evolution equations by the expression implied when the Friedmann constraint is solved for $y$ (for $y>0$). 
The center subspace is then the $1$-dimensional subspace of the tangent space at the equilibrium point given by $\Sigma_\pm=0$, $x-c_1/\sqrt{6}+\sqrt{2/3}\,\psi=0$. This means that $\psi$ is a regular local coordinate of the corresponding center manifold (which exists but is not necessarily unique according to the center manifold theory). The evolution equations yield that the center manifold can be approximated in terms of this local coordinate as
\[\Sigma_\pm=O(\psi^3),\quad x-c_1/\sqrt{6}+\sqrt{2/3}\,\psi=-\sqrt{\frac23}\frac{2c_1}{6-c_1^2}\psi^2+O(\psi^3),\quad\text{at $\psi=0$}.\]
Also higher orders in $\psi$ can be computed easily. We can therefore replace $x$ in the evolution equation \eqref{eq:evolutionpsipsi} by $\psi$ using this expansion and find a new evolution equation for $\psi$:
\[\psi'=-c_1 \psi^2+O(\psi^3).\]
It then follows that all solutions $\psi(\tau)$ of this differential equation approach the value $0$ of the equilibrium point, at least if the initial value of $\psi$ is sufficiently small. Hence, the center manifold is future stable. We compute that the solutions on the center manifold have the asymptotic expansions
\[\psi(\tau)=\frac1{c_1} \tau^{-1}+\phi_* \tau^{-2}+\ldots,\quad
x(\tau)=\frac{c_1}{\sqrt 6}-\sqrt{\frac23}\frac 1{c_1} \tau^{-1}+\ldots,\]
at $\tau\rightarrow\infty$, where $\phi_*$ is a free datum which also appears in higher terms in the expansion of $x(\tau)$.

\paragraph{Numerical studies.}
We have therefore demonstrated that the equilibrium point is future stable in the Bianchi I class. This is of course not sufficient to show that it is a future attractor. We shall now check the attractor property numerically, but provide no proof.
\begin{figure}[t]
  \begin{minipage}{0.49\linewidth}
    \centering
    \includegraphics[width=\textwidth]{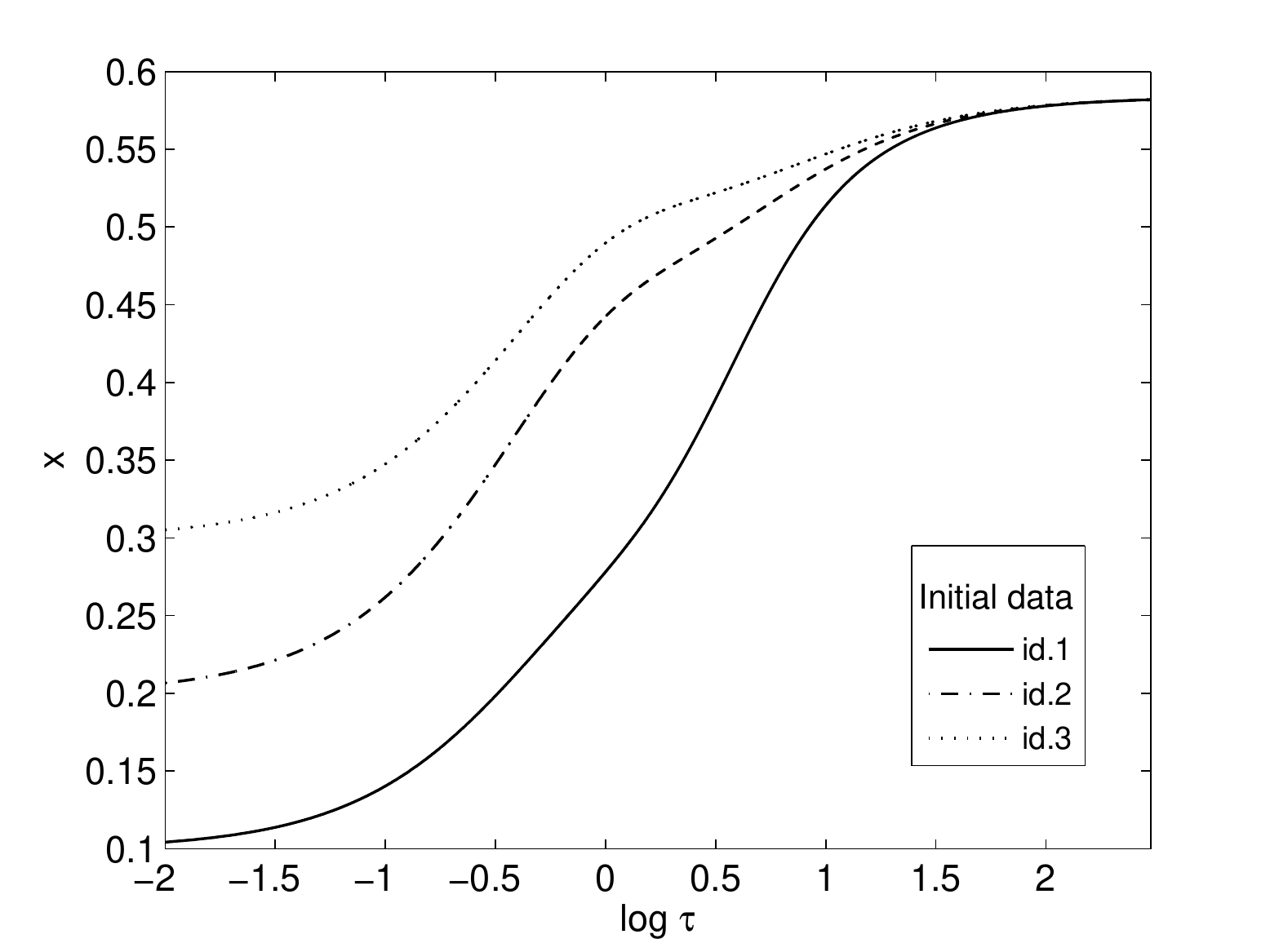}
    \caption{Scalar field kinetic energy $x$ (BI).}
    \label{fig:BI:x}
  \end{minipage}\hfill
  \begin{minipage}{0.49\linewidth}
    \centering
    \includegraphics[width=\textwidth]{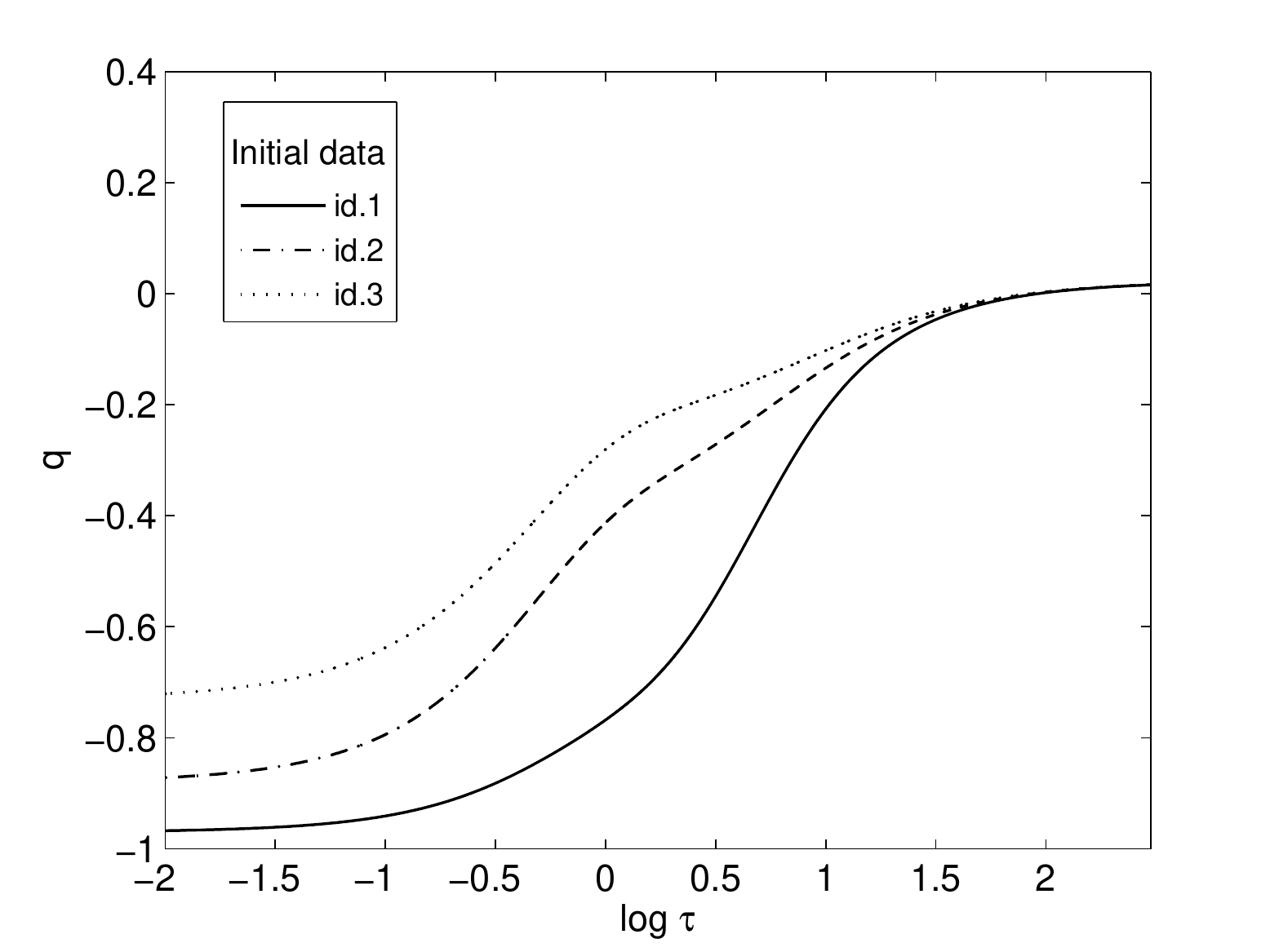}
    \caption{Deceleration scalar $q$ (BI). }
    \label{fig:BI:q}
  \end{minipage}
  \begin{minipage}{0.49\linewidth}
    \centering
    \includegraphics[width=\textwidth]{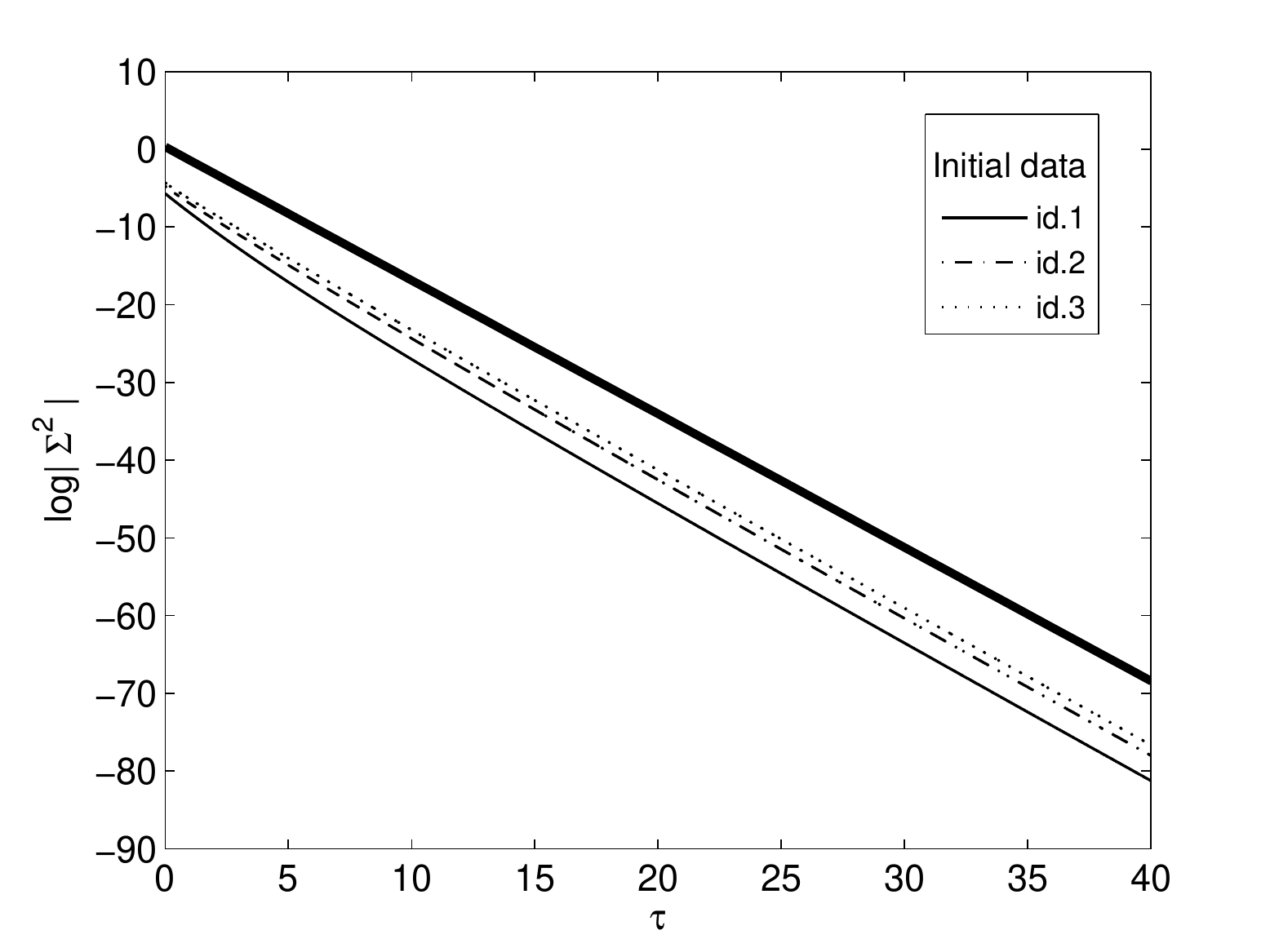}
    \caption{Decay of $\Sigma^2$ (BI). }
    \label{fig:BI:Sigma2}
  \end{minipage}\hfill
  \begin{minipage}{0.49\linewidth}
    \centering
    \includegraphics[width=\textwidth]{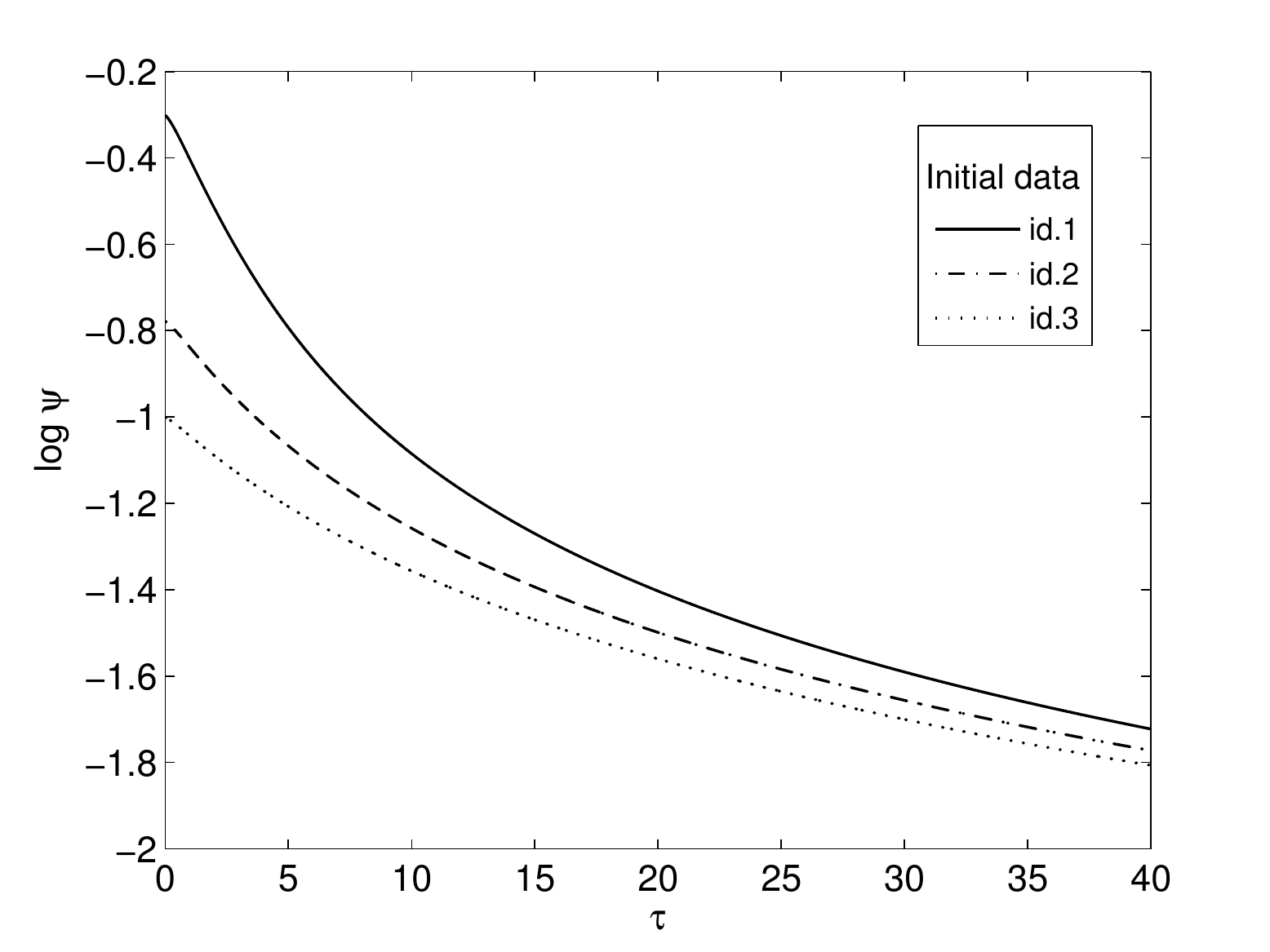}
    \caption{Scalar field $\psi=1/\phi$ (BI). }
    \label{fig:BI:phi}    
  \end{minipage}
\end{figure}
For the numerical computations, we choose $c_1=1.430$, $c_2=1.0$, which is in agreement with \Eqref{eq:Bianchirestriction}. Then, the equilibrium point is $\Sigma_{+}=\Sigma_{-}= 0 $, $x\approx 0.583$, $y\approx 0.811 $ and $q\approx 0.022$. In \Figsref{fig:BI:x} and \ref{fig:BI:q}, we demonstrate that the solutions corresponding to different inflationary data indeed approach these values asymptotically in the future. We point out that it is crucial here to use the modified evolution equations discussed in \Sectionref{sec:constraintdampingN} in order to get reliable numerical results; in particular close to the equilibrium point where there exists an unstable constraint violating mode. In any case, we have hence indeed constructed models numerically for which inflation lasts only for a finite time and a graceful exit into a permanent decelerated epoch follows. 
\Figref{fig:BI:Sigma2} shows that the numerical result is consistent with the exponential decay of the shear variables with exponent $(c_1^2-6)/2$ (the bold straight line in \Figref{fig:BI:Sigma2}) close to the equilibrium point as implied by the stability analysis above. Moreover, the $1/\tau$-decay of the function $\psi$ in \Figref{fig:BI:phi} as predicted from the center manifold analysis close to the equilibrium point is consistent.

\begin{figure}[t]
  \centering
  \includegraphics[width=0.49\textwidth]{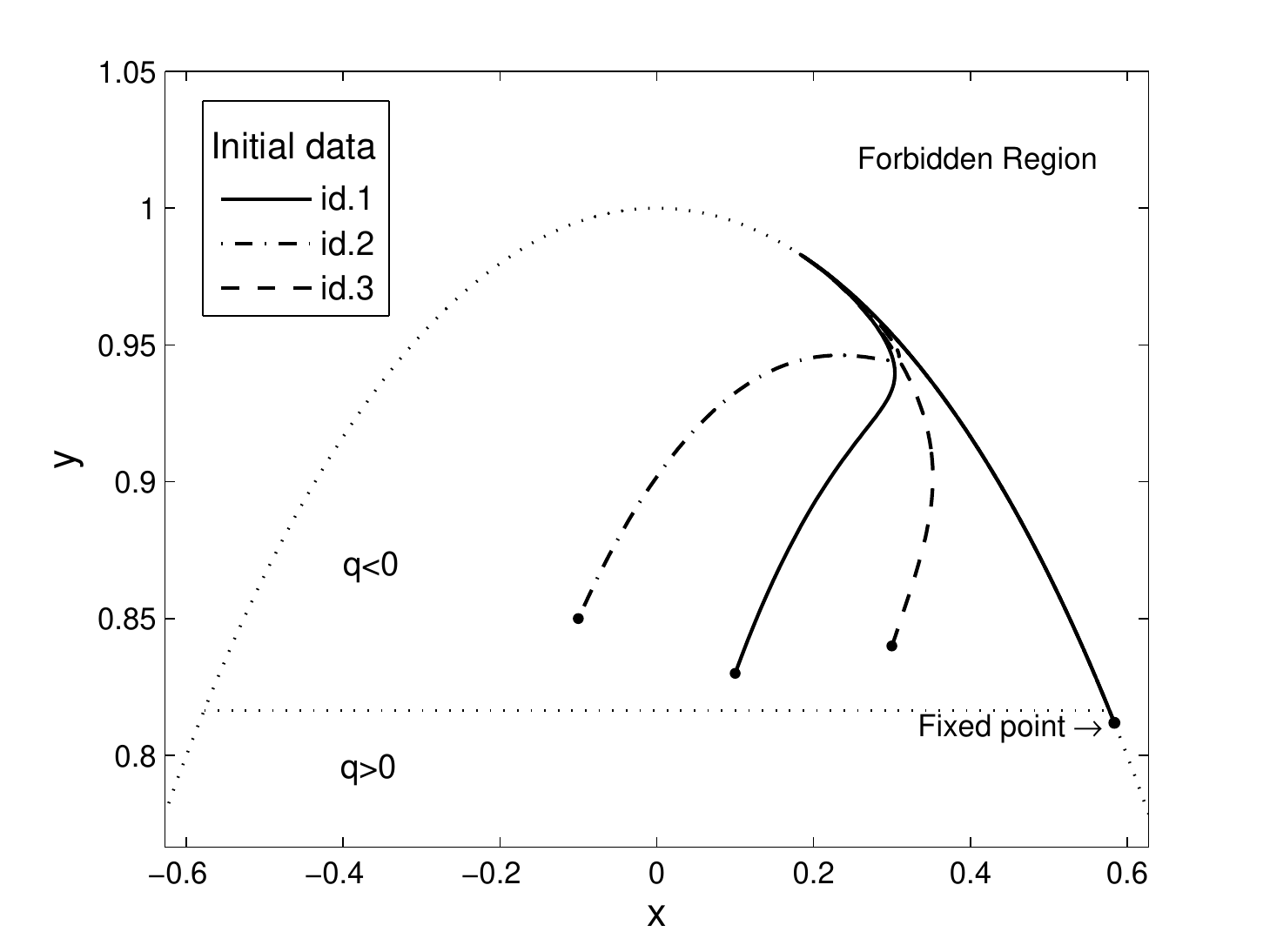}
  \caption{Projection of state space (BI).}
  \label{fig:BIxy}
\end{figure}
As a further representation of the dynamics in the Bianchi I case, consider \Figref{fig:BIxy}. In this figure, we plot the projection of the evolution orbits for three different initial data to the $x$-$y$-plane in the state space for the same case $c_1=1.43$, $c_1=1.0$ as before. According to the Friedmann constraint, the dynamics must take place in the interior of the circle $x^2+y^2=1$ (denoted by the dotted ellipsoidal curve in the figure), and \emph{on} this circle if and only if $\Sigma_\pm=0$ (i.e., the isotropic case). According to \Eqref{eq:q2}, the horizontal line $y=\sqrt{2/3}\approx 0.82$ in the figure corresponds to $q=0$, while everything above this line yields $q<0$ (inflation) and everything below is $q>0$ (deceleration). As we see, the three choices of initial data start in the inflationary regime (i.e., above the $q=0$ -- line). In accordance with the discussion of the quantity $S$ defined in \Eqref{eq:defR} in \Sectionref{sec:basicconsequencesN}, the orbits approach the circle $x^2+y^2=1$ rapidly and hence become isotropic very quickly during inflation. In fact, all orbits become close to the unit circle so quickly that some parts can hardly be distinguished in the figure. The orbits follow (as shown in the figure) the unit circle in such a way that eventually they all cross the $q=0$ -- line (graceful exit) and then approach the equilibrium point discussed before. 

\paragraph{Results.}
All this suggests that the equilibrium point above is indeed a future attractor for Bianchi I models in the decelerated regime if $\sqrt{2}<c_1<\sqrt{6}$. Hence if we start  with initial data in the inflationary regime then a graceful exit must occur under generic conditions. Since the equilibrium point is isotropic, anisotropies continue to decay even after inflation. We stress that only the limit $\lim_{\phi\rightarrow\infty}V'(\phi)/V(\phi)$ is relevant for the existence and main stability properties of the equilibrium point above. Hence it is conceivable that a similar dynamics can be found for more general classes of potentials. A particularly important example is the exponential potential studied in \cite{Kitada:1993wm}. For this potential only the $\phi^2$-term is missing in \Eqref{eq:potential}. In particular, the parameter $c_1$ has the same meaning, and one finds that the limit in \Eqref{eq:defFlimit} is the same. We therefore conjecture that generic inflationary Bianchi I with an exponential potential have a graceful exit after a finite time under the same conditions as here, namely if $\sqrt{2}<c_1<\sqrt{6}$. This would be remarkably consistent with the main result of \cite{Kitada:1993wm}, namely, that there is \textit{eternal} power-law inflation for $0<c_1<\sqrt{2}$. As a final remark, let us comment on the asymptotic behavior of the Hubble scalar $H$ for our models. Since $q$ approaches a stationary positive value asymptotically, $H$ decays towards zero exponentially according to \Eqref{eq:evolH}.

\subsubsection{Bianchi II \texorpdfstring{($N_1>0$, $N_2=N_3=0$)}{(N1>0,N2=N3=0)}}

\paragraph{Future attractors.}
In the Bianchi II case, we require the variable $N_1$ to be positive while $N_2$ and $N_3$ vanish. The first question to ask is whether the Bianchi I equilibrium point in the previous section is linearly stable within the class of Bianchi II solutions and hence is a conceivable future attractor for Bianchi II solution. However, it turns out  the subspace of the tangent space generated by $N_1$ is unstable (very similar to the perfect fluid case discussed in \cite{Wainwright:2005wss} where the Bianchi I equilibrium point $F$, which represents a flat FLRW solution, is unstable in Bianchi II).

We therefore construct now a genuinely Bianchi II equilibrium point with $y>0$ and $\psi=0$ as before. As in the Bianchi I case,  the condition $\Sigma_-'=0$ implies that $\Sigma_-=0$ since $S_-=0$, see \Eqref{eq:Sminus}. But, since $S_+$ does not necessary vanish, see \Eqref{eq:Splus}, the condition $\Sigma_+'=0$ yields that $\Sigma_+$ can be non-zero. The resulting algebraic system of equations enables us to find the following equilibrium point of the Bianchi II dynamical system
\begin{align*}
\Sigma_{+} &= \frac{2 \left(F_0^2-3\right)}{24+F_0^2},\quad \Sigma_{-}= 0,\quad N_{1}=\frac{6\sqrt{(12-F_0^2)(F_0^2-3)}}{24+F_0^2},\quad N_2=N_3=0,\\
 x &= -\frac{9 F_0}{24+F_0^2},\quad y= \frac{3\sqrt{12} \sqrt{12-F_0^2}}{24+F_0^2},\quad \psi= 0,
\end{align*}
for which
\[q=8\,\frac{F_0^2-3}{24+F_0^2}.
\] 
As before, $F_0$ can be expressed in terms of $c_1$ using \Eqref{eq:defFlimit}. The solution is real, and satisfies $y>0$ and $q>0$ if and only if 
\begin{equation}
\label{eq:BianchiIIFP}
\sqrt{2}< c_1 <\sqrt{8}.
\end{equation}
The same analysis as in the Bianchi I case reveals that this equilibrium point is future stable. Notice that this equilibrium point corresponds to the Collins-Stewart (II) perfect fluid solution $P_1^+(II)$ in \cite{Wainwright:2005wss} with $\Omega=18/(16+c_1^2)$ and $\gamma=6c_1^2/(16+c_1^2)$. \Eqref{eq:BianchiIIFP} yields the inequality $2/3<\gamma<2$.

\paragraph{Numerical studies.}
We claim that this equilibrium point is a future attractor for Bianchi II solutions, and support this claim by numerical studies.
\begin{figure}[t]
  \begin{minipage}{0.49\linewidth}
    \centering
    \includegraphics[width=\textwidth]{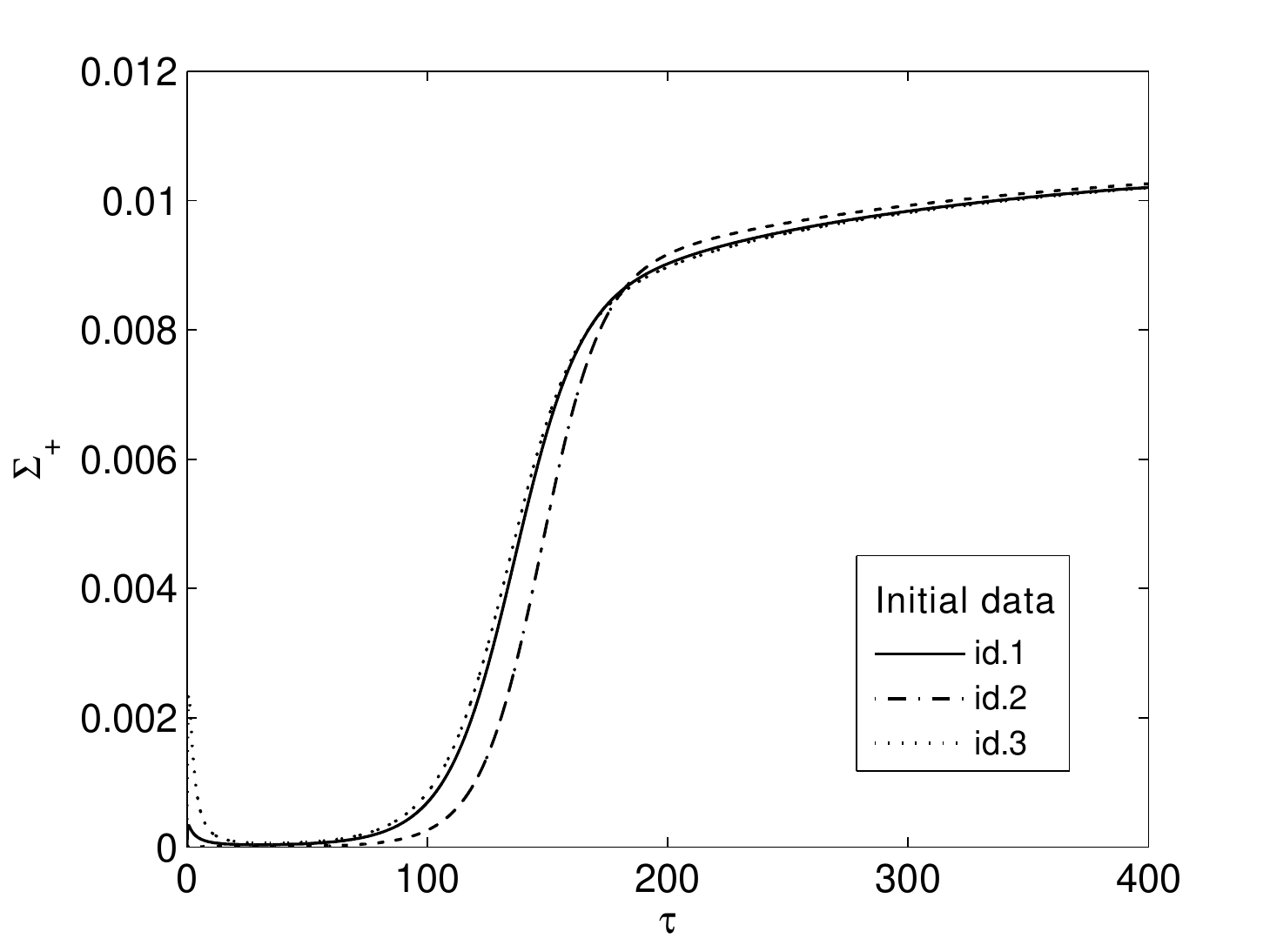}
    \caption{Anisotropy variable $\Sigma_{+}$ (BII). }
    \label{fig:BII:SigmaPlus}
  \end{minipage}
 \begin{minipage}{0.49\linewidth}
    \centering  
    \includegraphics[width=\textwidth]{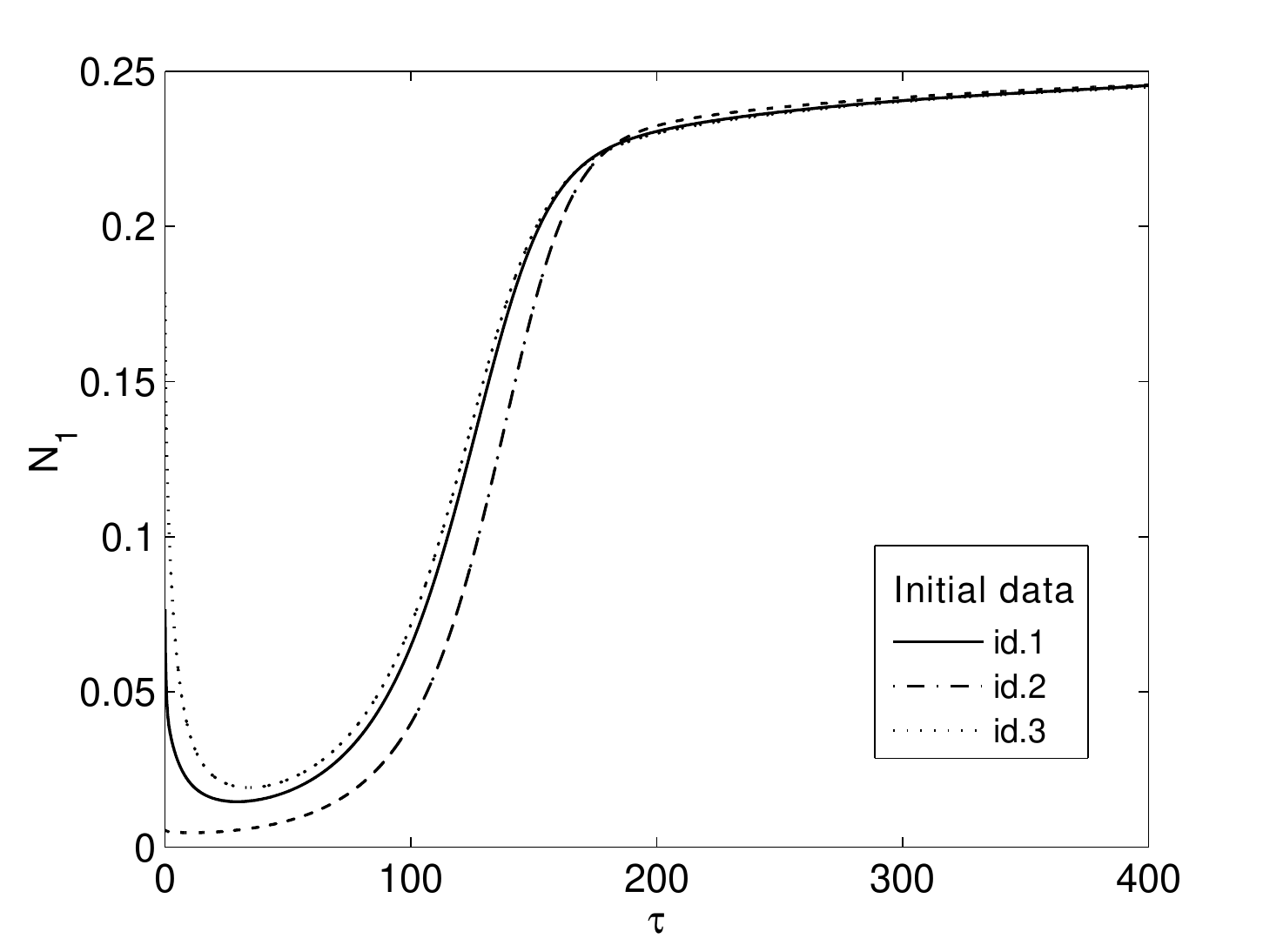}
    \caption{Quantity $N_1$ (BII). }
    \label{fig:BII:N1}
  \end{minipage}
  \begin{center}
    \begin{minipage}{0.49\linewidth}
      \centering
      \includegraphics[width=\textwidth]{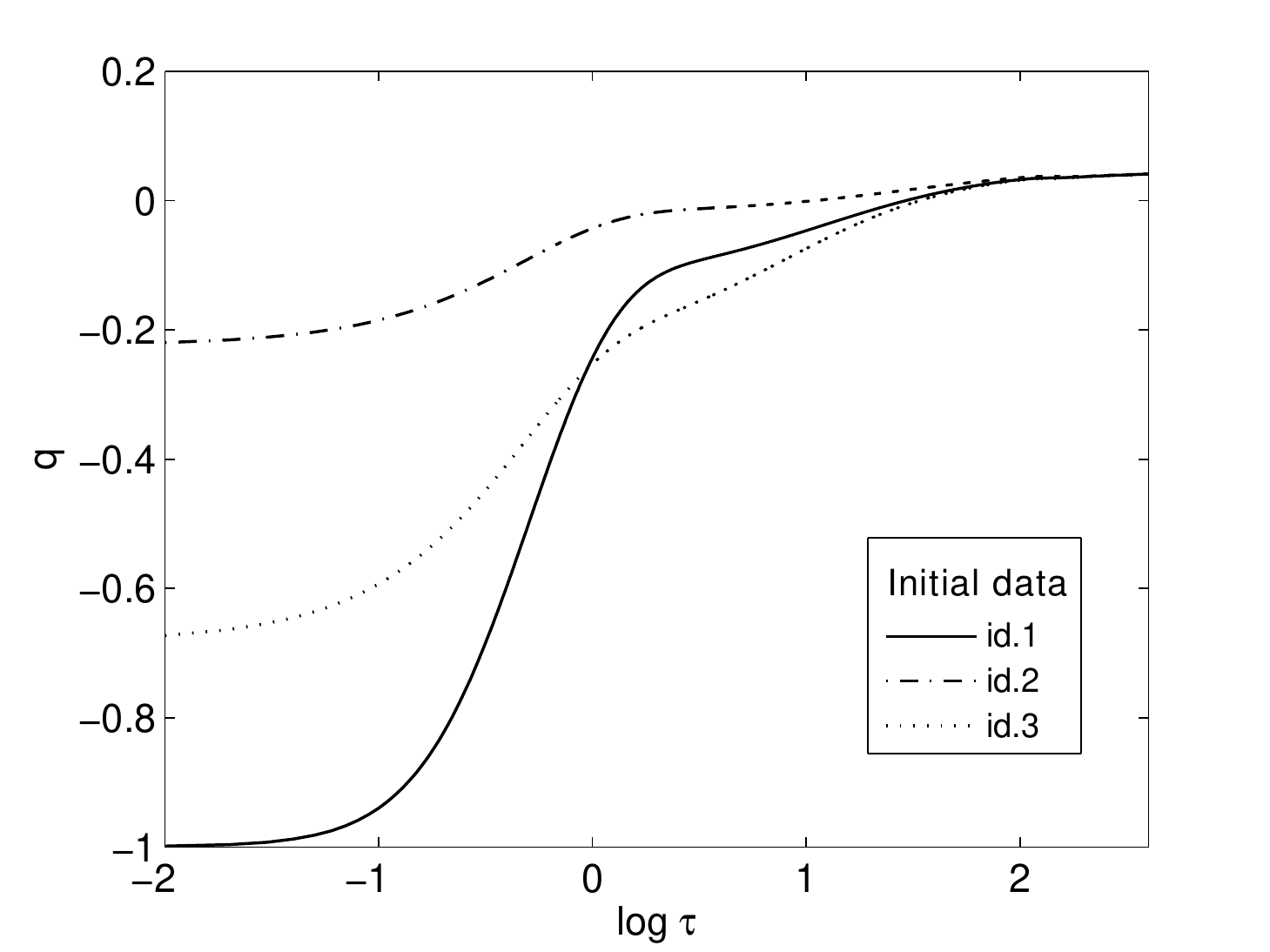}
      \caption{Deceleration scalar $q$ (BII).}
      \label{fig:BII:q}
    \end{minipage}
  \end{center}
\end{figure}
For this case, we  choose  $c_1=1.450$, $c_2=1.0$ which meets \Eqref{eq:BianchiIIFP}. The corresponding stationary values are $\Sigma_{+}\approx 0.011$, $\Sigma_{-}=0$, $N_1\approx 0.257 $,  $x\approx 0.588$, $y\approx 0.804$,  and  $q\approx 0.045$. In \Figsref{fig:BII:SigmaPlus}, \ref{fig:BII:N1} and \ref{fig:BII:q}, we demonstrate for various inflationary initial data that the evolution indeed approaches these values asymptotically. In particular, a graceful exit from inflation always happens. 

\paragraph{Results.}
We have therefore identified a future attractor of type Bianchi II in the decelerated regime if $\sqrt{2}<c_1<\sqrt{8}$. Hence under this condition, generic initially inflationary solutions have a graceful exit from inflation very similar to the Bianchi I case. The main difference to the Bianchi I case is the asymptotic behavior of the spatial curvature and anisotropy variables. Because the values for $\Sigma_+$ and $N_1$ of the attractor are non-zero, isotropization and decay of spatial curvature stops at the time of the graceful exit. Even more so, as we see in \Figsref{fig:BII:SigmaPlus} and \ref{fig:BII:N1}, these quantities can grow again rapidly after inflation.
In particular, this shows that one of the basic cornerstones of the standard model of cosmology, namely isotropy, may in general not  be explained satisfactorily by a finite phase of inflation only. As our example here shows, it may in fact be possible that anisotropies and spatial curvature, even if they are extremely small by the end of inflation, grow again when inflation is over. Hence, it may be necessary to introduce additional so far unknown mechanisms.

\subsubsection{Bianchi \texorpdfstring{VI$_0$ ($N_1=0$, $N_2>0$, $N_3<0$)}{VI0 (N1=0, N2>0, N3>0)}}
\paragraph{Future attractors.}
For the Bianchi VI$_0$ case, we assume $N_1=0$, and that $N_2$ and $N_3$ have opposite signs. Again the Bianchi I equilibrium point above could be a future asymptotic end state for general Bianchi VI$_0$ orbits. In the same way as in Bianchi II, however, it is future unstable in Bianchi VI$_0$. The Bianchi II equilibrium point above could also be a future asymptotic end state (when a rotation is applied to the orthonormal frame). However, also this is unstable in Bianchi VI$_0$.

We therefore attempt to construct a Bianchi VI$_0$ equilibrium point with $q>0$ and $y>0$ as before. 
The condition $N_2'=N_3'=0$, but $N_2,N_3\not=0$, implies that $q=-2\Sigma_+$ and $\Sigma_-=0$, see \Eqref{eq:evolutioneqN2} and \eqref{eq:evolutioneqN3}. Then \Eqref{eq:evolutioneqSigma} together with \Eqref{eq:Sminus} yields that $N_2=-N_3$. Using these relations, we find that following result
\begin{align*}
&\Sigma_{+} = \frac{3-F_0^2}{6+F_0^2}, &
&\Sigma_{-}= 0, &
&N_{1}=0,\quad
N_2=-N_3=\frac{3\sqrt{3}\sqrt{F_0^2-3}}{6+F_0^2}, & \\ 
&x = -\frac{3 F_0}{6+F_0^2}, &
&y= \frac{3 \sqrt{6}}{6+F_0^2}, &
&\psi= 0,&
\end{align*}
where
\[q=2\frac{F_0^2-3}{F_0^2+6}.
\]
Expressing $F_0$ in terms of $c_1$ as before, we find that the only restriction is
\begin{equation}\label{eq:BianchiVIFP}
c_1>\sqrt{2}.
\end{equation}
Notice that this equilibrium point corresponds to the Collins (VI$_0$) perfect fluid solution with fixed point $P_1^+(VI_0)$ in \cite{Wainwright:2005wss} with $\Omega=6/(4+c_1^2)$ and $\gamma=2c_1^2/(4+c_1^2)$. The inequality for $c_1$ again yields the inequality $2/3<\gamma<2$.

\paragraph{Results.}
Since the results are so similar to the Bianchi II case, except that in addition here the modulus of $N_2$ and $N_3$ approaches the same value asymptotically while $N_1=0$, we do not show any numerical results now. We come to the same conclusions as in the Bianchi II case that for generic inflationary Bianchi VI$_0$ initial data there is a graceful exit if $c_1>\sqrt{2}$. Again the future attractor is neither isotropic nor spatially flat.

\subsubsection{Bianchi \texorpdfstring{VII$_0$ ($N_1=0$, $N_2,N_3>0$)}{VII0 (N1=0, N2,N3>0)}}
We proceed with the Bianchi VII$_0$ case where $N_1$ is assumed to vanish and $N_2$ and $N_3$ are both positive. At a first glance, the situation seems to be similar to the Bianchi VI$_0$ case. However, as $N_2$ and $N_3$ have the same sign now, the Friedmann constraint does not imply boundedness of $N_2$ and $N_3$.

\paragraph{Future attractors.}
The numerical studies which we present in more detail below suggest that for generic inflationary Bianchi VII$_0$ initial data, $N_2$ and $N_3$ are unbounded. This means that we must not look for equilibrium points in order to describe the future asymptotics. However, as we see below, the numerics suggest that
all quantities $\Sigma_\pm$, $x$, $y$ and $\psi$ approach stationary values (as it was the case for all previous Bianchi cases), namely that
\begin{equation}
  \label{eq:BVIIattractor}
  \Sigma_\pm\rightarrow 0, \quad x\rightarrow x_*,\quad  y\rightarrow y_*,\quad 
  \psi\rightarrow 0,\quad \text{for $\tau\rightarrow\infty$},
\end{equation}
where the values $x_*$ and $y_*$ are universal (i.e., they only depend on $c_1$ and not on the initial data). We claim now that for every choice of the parameters $c_1$ and $c_2$ (in a certain range), we can construct a future attractor for Bianchi VII$_0$ solutions in the decelerated regime as follows.
 According to \Eqsref{eq:q} and \eqref{eq:BVIIattractor}, the deceleration scalar $q$ must approach a universal value $q_*=2x_*^2-y_*^2$.
\Eqsref{eq:evolutioneqN2} and \eqref{eq:evolutioneqN3} then imply that
\begin{equation}
  \label{eq:BVIIattractorN}
  N_2(\tau)\rightarrow N_{2*} e^{q_*\tau},\quad N_3(\tau)\rightarrow N_{3*} e^{q_*\tau},
\end{equation}
for large $\tau$, where $N_{2*}$ and $N_{3*}$ are some positive constants. On the other hand, \Eqsref{eq:evolutioneqSigma} with $\Sigma_\pm\rightarrow 0$ and $\Sigma_{\pm}'\rightarrow 0$ imply that $S_+,S_-\rightarrow 0$. According to \Eqsref{eq:Splus} and \eqref{eq:Sminus} these quantities read in the Bianchi VII$_0$ case:
\begin{equation}
  \label{eq:SPMBVII}
  S_+=\frac 16(N_2-N_3)^2,\quad S_-=\frac 1{2\sqrt 3}(N_2^2-N_3^2).
\end{equation}
This implies in particular that $N_{2*}=N_{3*}$ (but also gives a restriction on the next order of the asymptotic expansion for $N_2$ and $N_3$ which we do not discuss here in detail now). The quantity $K$, which according to \Eqref{eq:K} is
$K=(N_2-N_3)^2/2$
in the Bianchi VII$_0$ case,
must also approach zero. We can therefore  consider \Eqref{eq:hamiltonianconstraint}, \eqref{eq:evolutioneqx} and \eqref{eq:evolutioneqy} easily in the limit $\tau\rightarrow\infty$, and find algebraic equations for $x_*$ and $y_*$ which are the same as in the Bianchi I case. The solution is
\begin{equation}
  \label{eq:BVIIattractorstationvalues}
  x_*=\frac{c_1}{\sqrt 6},\quad y_*=\sqrt{1-\frac{c_1^2}6},\quad q_*=\frac{c_1^2-2}2.
\end{equation}
In particular, if 
\begin{equation}
  \label{eq:c1BVII}
  \sqrt{2}<c_1< \sqrt{6},
\end{equation}
we claim that generic initially inflationary Bianchi VII$_0$ solutions approach the asymptotic behavior given by \Eqsref{eq:BVIIattractor}, \eqref{eq:BVIIattractorN} with $N_{2*}=N_{3*}$ and \eqref{eq:BVIIattractorstationvalues} in the future.

\paragraph{Numerical studies.}
\begin{figure}[t]
  \begin{minipage}{0.49\linewidth}
    \centering
    \includegraphics[width=\textwidth]{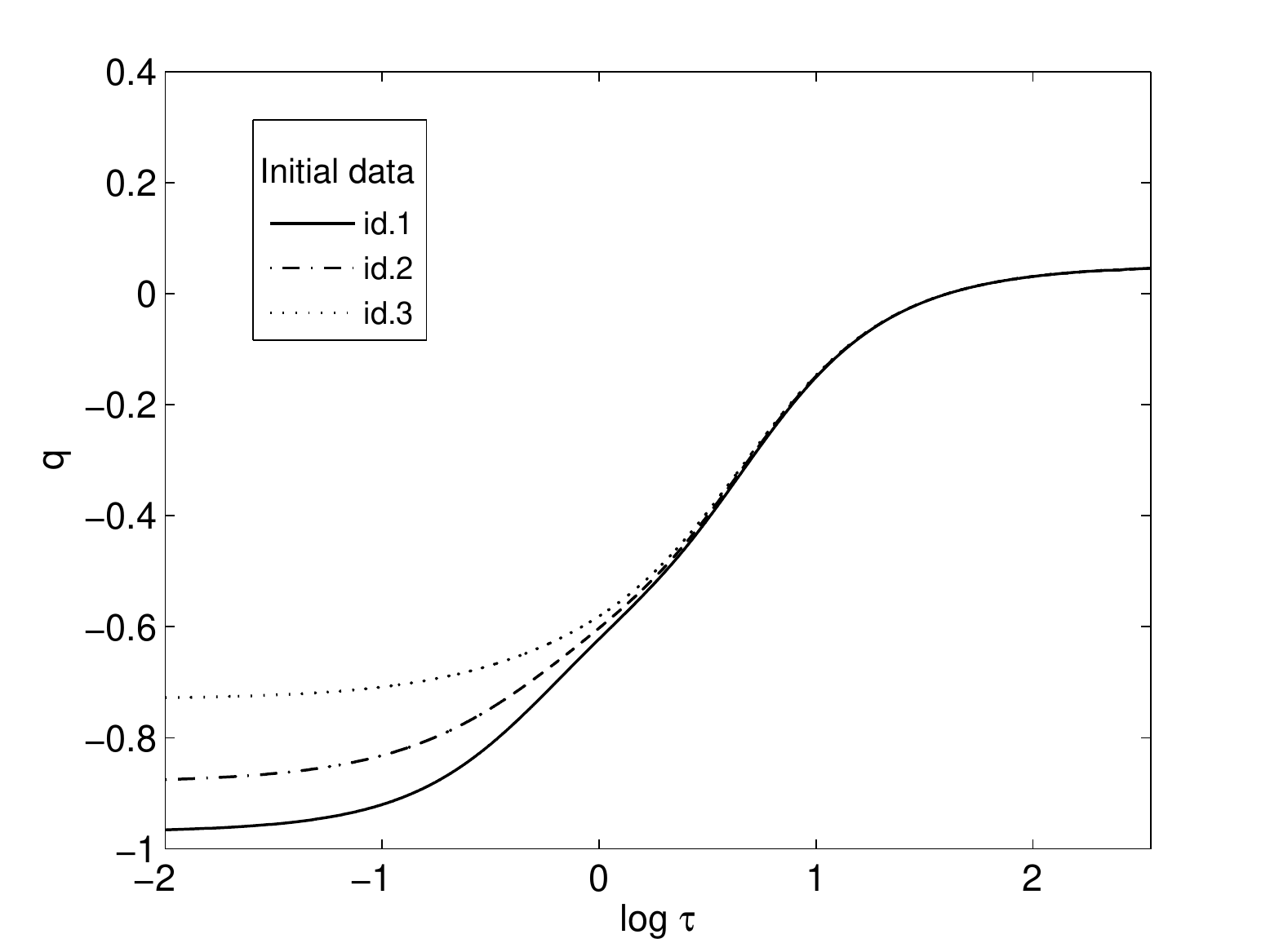}
    \caption{Deceleration scalar $q$,  (BVII$_0$). }
    \label{fig:BVII:q}
  \end{minipage}\hfill
  \begin{minipage}{0.49\linewidth}
    \centering
    \includegraphics[width=\textwidth]{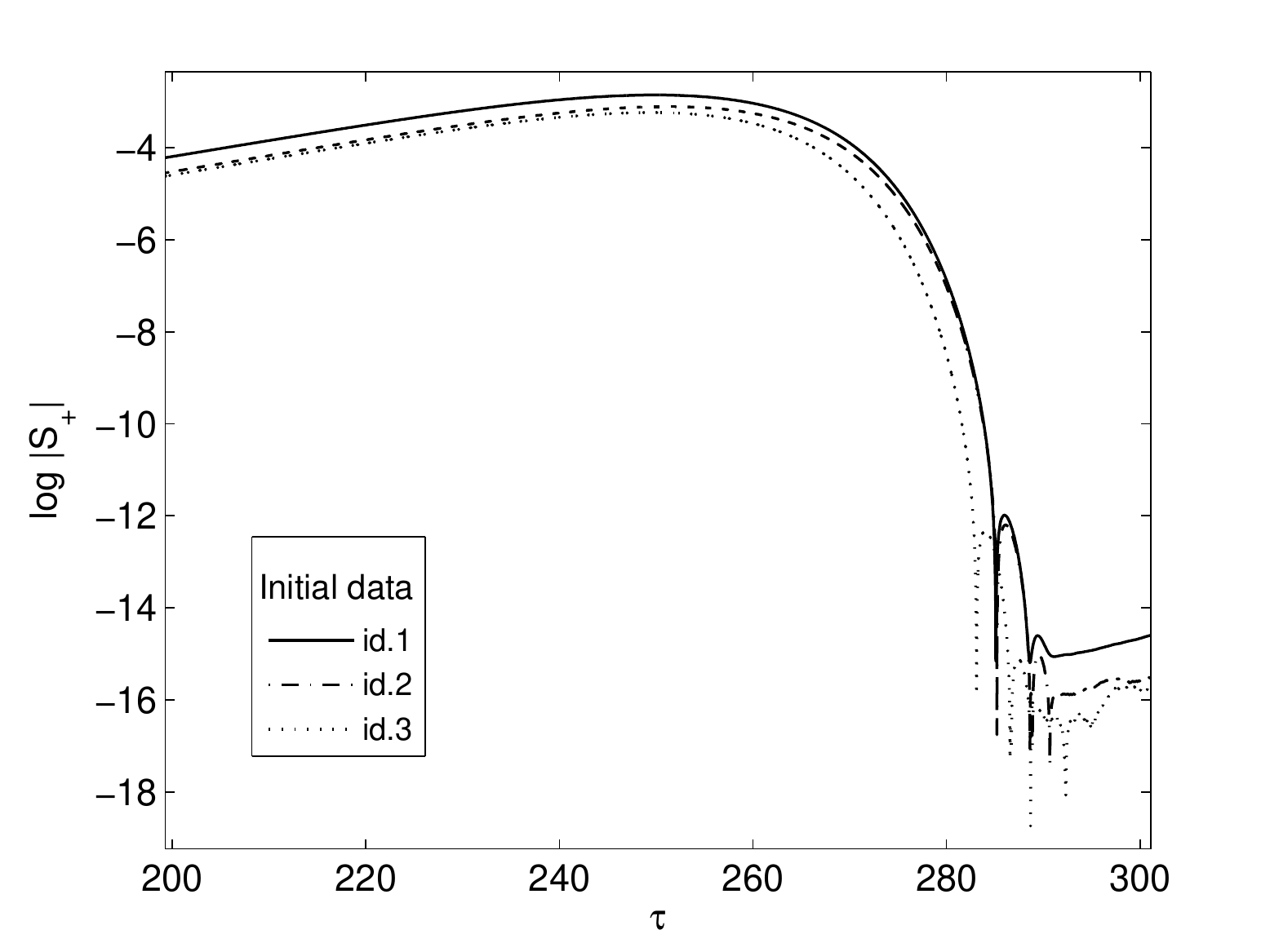}
    \caption{Oscillatory decay of $|S_+|$ (BVII$_0$). }
    \label{fig:BVII:SPlus}
  \end{minipage}
  \begin{minipage}{0.49\linewidth}
    \centering
    \includegraphics[width=\textwidth]{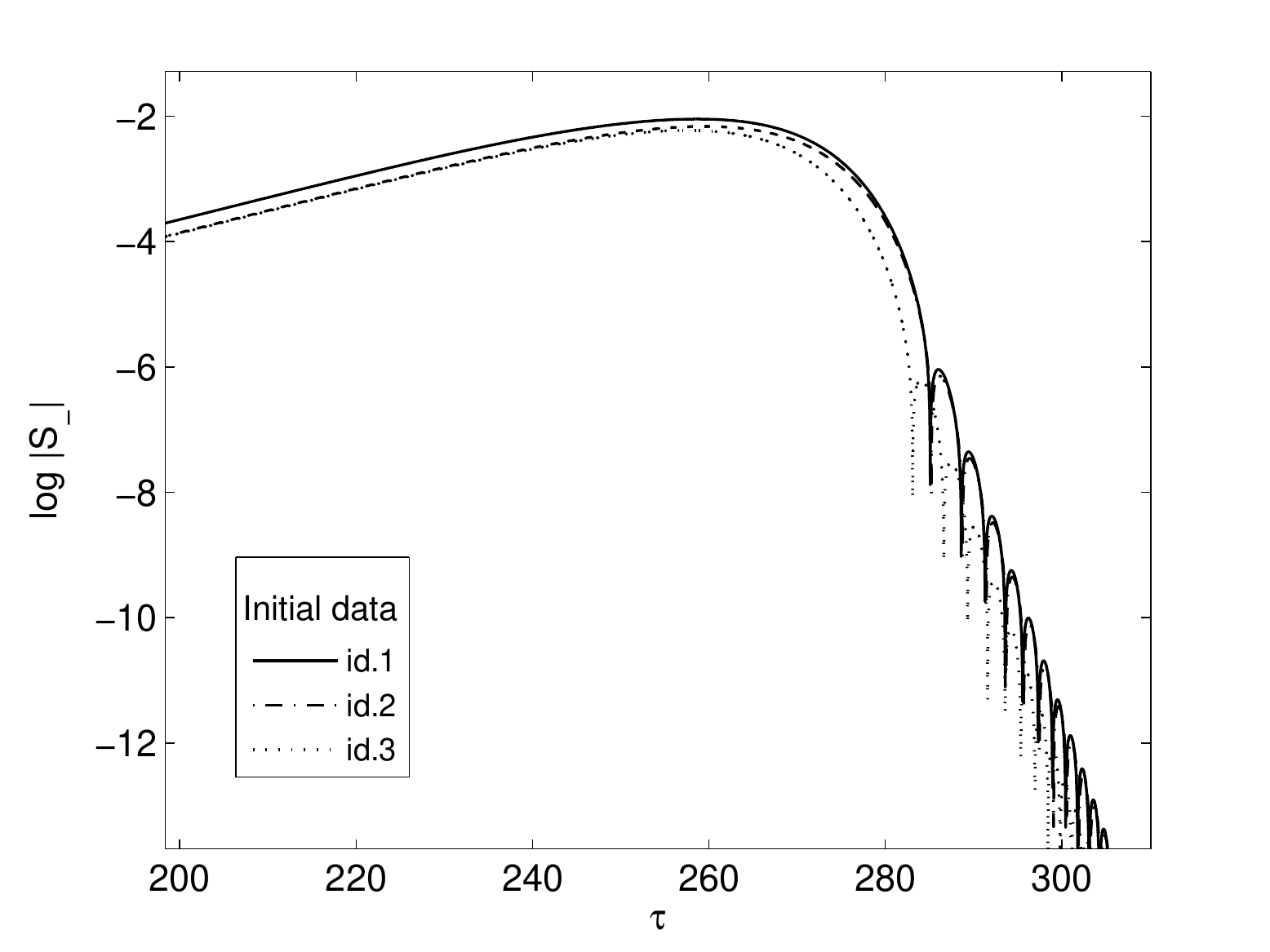}
    \caption{Oscillatory decay of $|S_-|$ (BVII$_0$). }
    \label{fig:BVII:SMinus}
  \end{minipage}\hfill
  \begin{minipage}{0.49\linewidth}
    \centering \includegraphics[width=\textwidth]{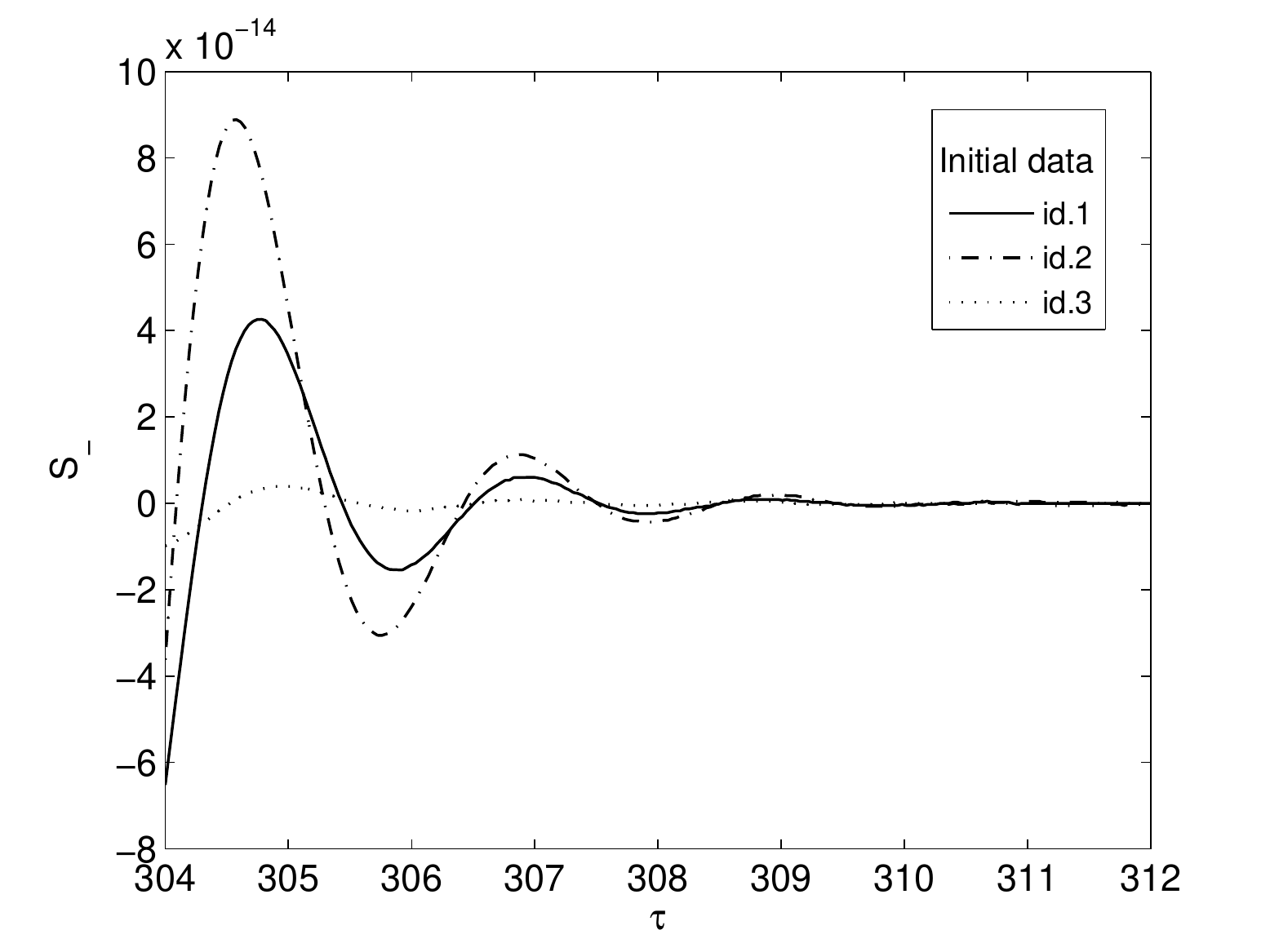}
    \caption{Late time oscillations of $S_-$ (BVII$_0$). }
    \label{fig:BVII:SMinusOscillations}
  \end{minipage}
\end{figure}

\begin{figure}[t]
  \begin{minipage}{0.49\linewidth}
    \centering 
    \includegraphics[width=\textwidth]{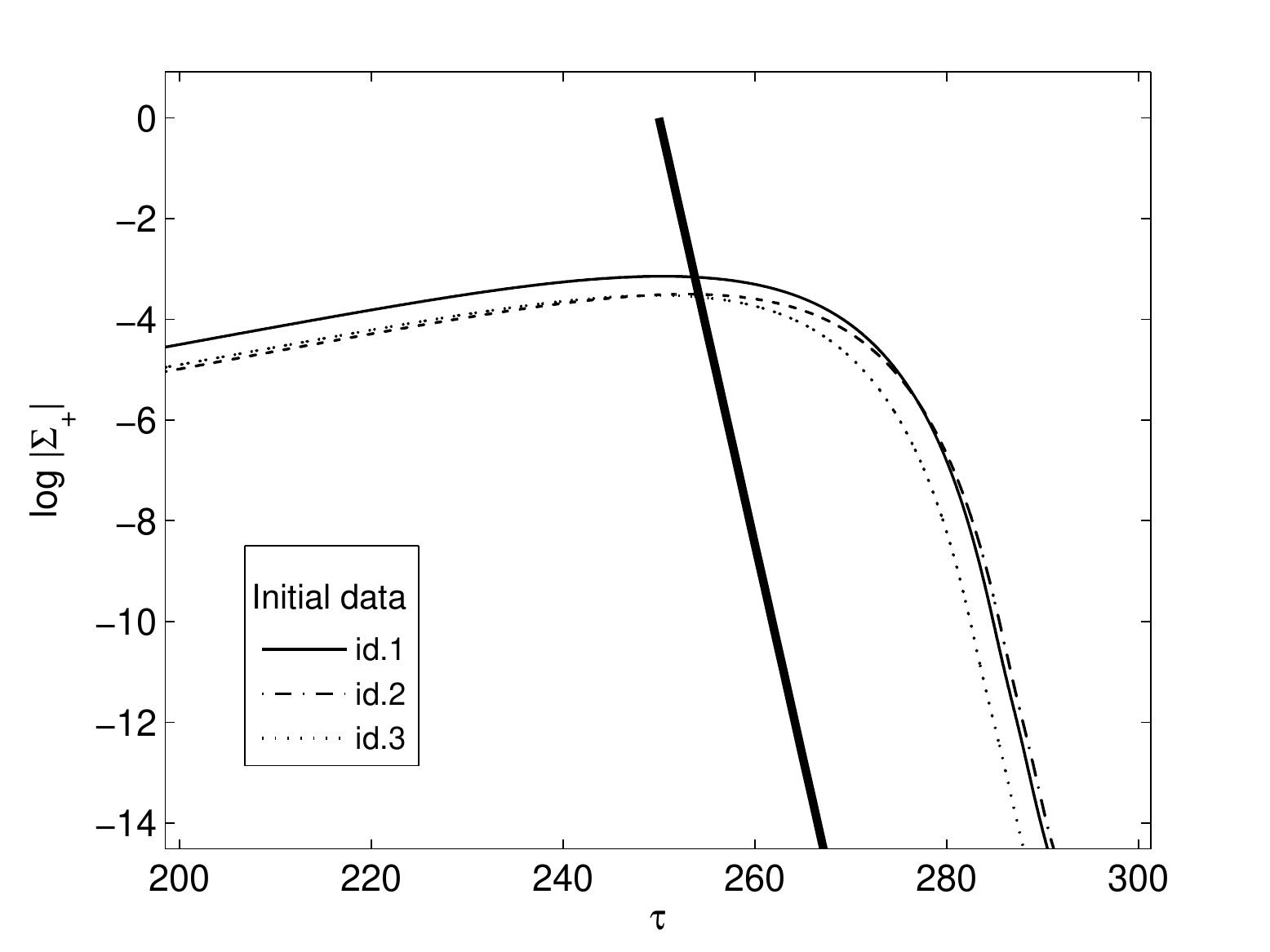}
    \caption{Decay of $\Sigma_+$ (BVII$_0$). }
    \label{fig:BVII:SigmaPlus}
  \end{minipage}
  \begin{minipage}{0.49\linewidth}
    \centering 
    \includegraphics[width=\textwidth]{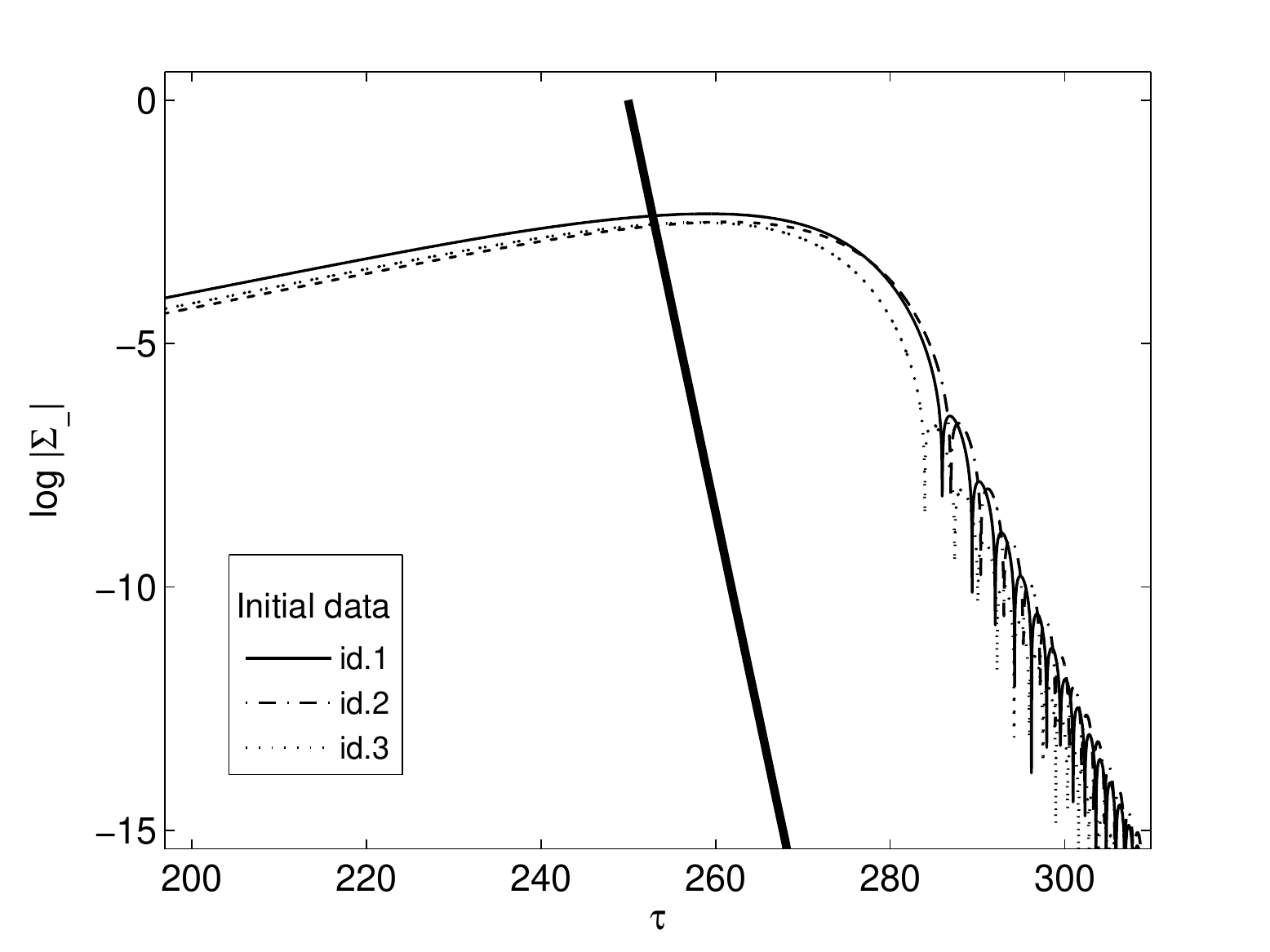}
    \caption{Oscillatory decay of $\Sigma_-$ (BVII$_0$). }
    \label{fig:BVII:SigmaMinus}
  \end{minipage}
\end{figure}

For our numerical studies, we choose $c_{1}= 1.45$ and $c_2=1.0$. This implies that $x_*\approx 0.59$, $y_*\approx 0.81$ and $q_*\approx 0.05$ according to \Eqref{eq:BVIIattractorstationvalues}. Indeed, the variables $\Sigma_\pm$, $x$, $y$ and $\psi$ approach these stationary values generically, and hence in particular $q$ (see \Figref{fig:BVII:q}). We therefore find that our initially inflationary Bianchi VII$_0$ solutions have graceful exits from inflation. Of major interest for the Bianchi VII$_0$ case is the behavior of the variables $N_2$ and $N_3$ which are supposed to behave like \Eqref{eq:BVIIattractorN} with $N_{2*}=N_{3*}$ so that $S_+,S_-\rightarrow 0$. In \Figsref{fig:BVII:SPlus} and \ref{fig:BVII:SMinus} we see that this decay towards zero indeed happens exponentially. However, the decay is not monotonic and the frequency of the oscillations increases as time progresses. While $S_+$ is non-negative, the quantity $S_-$ can have any sign, cf.\ \Figref{fig:BVII:SMinusOscillations}. The evolution equations for $\Sigma_\pm$ therefore suggest that also $\Sigma_\pm$ should be oscillatory.  The plots in \Figsref{fig:BVII:SigmaPlus} and \ref{fig:BVII:SigmaMinus} confirm that while $\Sigma_+$ decays monotonically with the same rate as in the Bianchi I case (represented by the bold line), the quantity $\Sigma_-$ decays with a smaller rate than in the Bianchi I case and is oscillatory with increasing frequency.

\paragraph{Results.}
The Bianchi VII$_0$ case differs significantly from the previous Bianchi cases due to the unboundedness of the quantities $N_2$ and $N_3$. It is particularly interesting to note that the situation is not the same in pure vacuum where $N_2$ and $N_3$ have been proven to be bounded for generic Bianchi VII$_0$ solutions; see~Theorem 1.1 in \cite{Ringstrom:2001es}. In our case here, $N_2$ and $N_3$ grow exponentially as in \Eqref{eq:BVIIattractorN} where $q_*$ is given in \Eqref{eq:BVIIattractorstationvalues} such that $S_+$ and $S_-$ given by \Eqref{eq:SPMBVII} go to zero. This leads to a subtle oscillatory behavior which is passed on as oscillations to the otherwise exponentially decaying shear variables $\Sigma_\pm$. This dynamics should be studied in more detail in future work, in particular, the dependency of the oscillation frequency on time and the decay rate of $\Sigma_-$ in \Figref{fig:BVII:SigmaMinus}. Nevertheless, our main claim that  generic  initially inflationary Bianchi VII$_0$ solutions have a graceful exit from inflation (if $c_1$ is chosen appropriately), has been confirmed by our numerical investigations. This suggests that the variables $\Sigma_\pm$, $x$, $y$ and $\psi$ approach stationary values similar to previous Bianchi cases. Since the future attractor is isotropic in the Bianchi VII$_0$ case, anisotropies (represented by $\Sigma_\pm$) and the spatial curvature (represented by the variables $S_+$, $S_-$ and $K$) continue to decay even after inflation.

\subsubsection{Bianchi VIII \texorpdfstring{($N_1<0$, $N_2,N_3>0$)}{(N1<0, N2,N3>0)}}

The most complicated case, which we study in this paper, is Bianchi VIII where we have the negative variable $N_1$ in addition to the positive variables $N_2$ and $N_3$. Similar to Bianchi VII$_0$, the quantities $N_2$ and $N_3$ are not bounded by the Friedmann constraint, but $\Sigma_\pm$, $x$, $y$ and $N_1$ are.

\paragraph{Future attractors.}
The numerical studies presented below suggest that all quantities $\Sigma_\pm$, $x$, $y$, $N_1$ and $\psi$ approach stationary values during the evolution, but that $N_2$ and $N_3$ are unbounded. Hence the situation appears to be quite similar to the Bianchi VII$_0$ case. The numerical computations suggest that
\begin{equation}
  \label{eq:BVIIIattractor}
  \Sigma_+\rightarrow\Sigma_{+*},\quad\Sigma_-\rightarrow 0,\quad N_1\rightarrow 0,\quad x\rightarrow x_*,\quad  y\rightarrow y_*,\quad 
  \psi\rightarrow 0,\quad \text{for $\tau\rightarrow\infty$},
\end{equation}
where the values $x_*$, $y_*$ and $\Sigma_{+*}$ are universal and only depend on the parameter $c_1$ of the potential. We notice that in particular $\Sigma_+$ does in general not approach zero in the Bianchi VIII case (in contrast to Bianchi VII$_0$).
 According to \Eqref{eq:q}, the deceleration scalar $q$ approaches the universal value 
\begin{equation}
  \label{eq:qBVIII}
  q_*=2\Sigma_{+*}^2+2x_*^2-y_*^2.
\end{equation}
\Eqsref{eq:evolutioneqN1} -- \eqref{eq:evolutioneqN3} imply that
\begin{equation}
  \label{eq:BVIIIattractorN}
  N_1(\tau)\rightarrow N_{1*} e^{(q_*-4\Sigma_{+*})\tau},\quad N_2(\tau)\rightarrow N_{2*} e^{(q_*+2\Sigma_{+*})\tau},\quad N_3(\tau)\rightarrow N_{3*} e^{(q_*+2\Sigma_{+*})\tau},
\end{equation}
for large $\tau$ with constants $N_{1*}<0$, $N_{2*}>0$ and $N_{3*}>0$. \Eqsref{eq:Splus} and \eqref{eq:Sminus} read 
\begin{align*}
S_+&=\frac 16\left((N_2-N_3)^2-2N_1^2-|N_1|(N_2+N_3)\right),\\
S_-&=\frac 1{2\sqrt 3}\left((N_2^2-N_3^2)+|N_1|(N_2-N_3)\right).
\end{align*}
Since \Eqref{eq:evolutioneqSigma} with $\Sigma_-\rightarrow 0$ and $\Sigma_{-}'\rightarrow 0$ implies that $S_-\rightarrow 0$ and since $N_1\rightarrow 0$, it follows in the same way as in Bianchi VII$_0$ that $N_2-N_3\rightarrow 0$ and $N_2^2-N_3^2\rightarrow 0$. This means in particular that $N_{2*}=N_{3*}$. \Eqref{eq:evolutioneqSigma} with $\Sigma_+\rightarrow \Sigma_{+*}\not=0$ and $\Sigma_{+}'\rightarrow 0$, however, implies that $S_+$ must not vanish asymptotically, but instead
\begin{equation}
  \label{eq:BVIIIattractorN1}
  \frac 16|N_1|(N_2+N_3)\rightarrow (2-q_*)\Sigma_{+*}.
\end{equation}
A necessary condition for $|N_1|(N_2+N_3)$ approaching a finite non-zero value is that the exponents of $N_1$, $N_2$ and $N_3$ in \Eqref{eq:BVIIIattractorN} match:
\begin{equation}
  \label{eq:matchexponents}
  q_*-4\Sigma_{+*}+q_*+2\Sigma_{+*}=0\quad\Leftrightarrow\quad q_*=\Sigma_{+*}.
\end{equation}
Since we can write $K$ according to \Eqref{eq:K} as 
\[K=\frac1{12}\left((N_2-N_3)^2+2|N_1|(N_2+N_3)\right),\]
and hence conclude that $K\rightarrow (2-q_*)\Sigma_{+*}$, the Friedmann constraint implies
\begin{equation}
  \label{eq:constrBVIII}
  1=\Sigma_{+*}^2+x_*^2+y_*^2+(2-q_*)\Sigma_{+*}.
\end{equation}
We get another equation from \Eqref{eq:evolutioneqx} with $x'=0$, which yields
\begin{equation}
  \label{eq:xevolBVIII}
  x_*(q_*-2)-F_0 y_*^2=0.
\end{equation}
We now solve the algebraic equations \eqref{eq:matchexponents}, \eqref{eq:constrBVIII}, \eqref{eq:xevolBVIII} for the unknowns $\Sigma_{+*}$, $x_*$ and $y_*$ where we replace $q_*$ by \Eqref{eq:qBVIII}. Under the condition that $y_*>0$ and that also \Eqref{eq:evolutioneqy} with $y'=0$ must be satisfied, we find precisely one solution
\begin{equation}
  \label{eq:BVIIIattractorstatval}
\Sigma_{+*}
=\frac{c_1^2-2}{2\left(c_1^2+1\right)},\quad 
x_*=\sqrt{\frac{3}{2}}\,\frac{c_1}{c_1^2+1}, \quad
y_*=\sqrt{\frac{3}{2}}\,\frac{\sqrt{c_1^2+2}}{c_1^2+1}.
\end{equation}
This implies that
\[q_*=\frac{c_1^2-2}{2(c_1^2+1)},\]
and hence this represents a decelerated epoch if and only if
\[c_1>\sqrt{2}.\]
As an additional check, we compute the exponent of $N_1$ in \Eqref{eq:BVIIIattractorN} in order to confirm that it is indeed negative
\[q_*-4\Sigma_{+*}=\frac32\,\frac{2-c_1^2}{1+c_1^2}.\]
Hence, 
we claim that if $c_1>\sqrt{2}$, generic Bianchi VIII solutions behave like \Eqsref{eq:BVIIIattractor}, \eqref{eq:BVIIIattractorN} with $N_{2*}=N_{3*}$, \eqref{eq:BVIIIattractorN1} and \eqref{eq:BVIIIattractorstatval} asymptotically for $\tau\rightarrow\infty$.

\paragraph{Numerical studies.}
\begin{figure}[t]
  \begin{minipage}{0.49\linewidth}
    \centering
    \includegraphics[width=\textwidth]{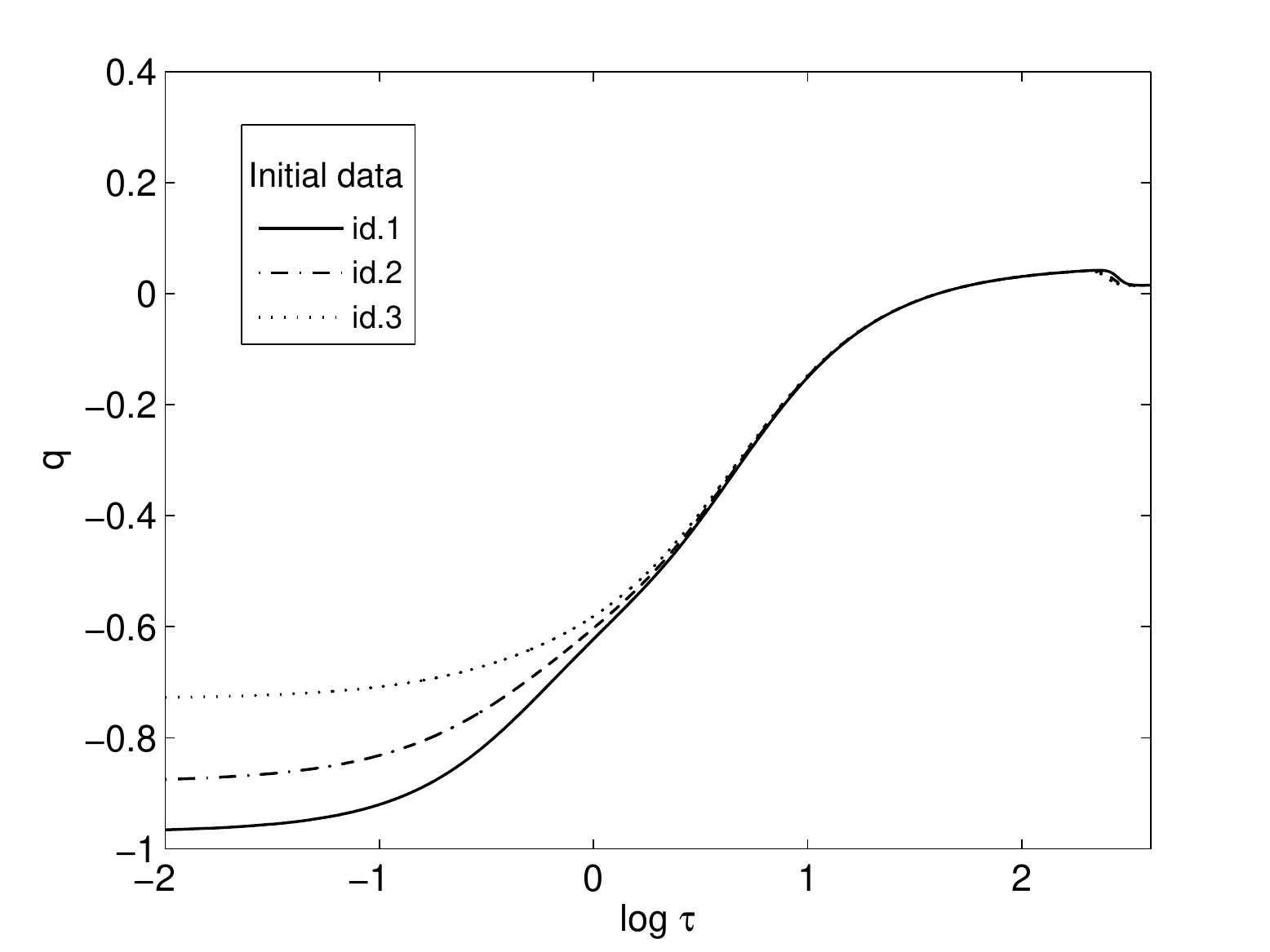}
    \caption{Deceleration scalar $q$ (BVIII). }
    \label{fig:BVIII:q}
  \end{minipage}\hfill  \begin{minipage}{0.49\linewidth}
    \centering
    \includegraphics[width=\textwidth]{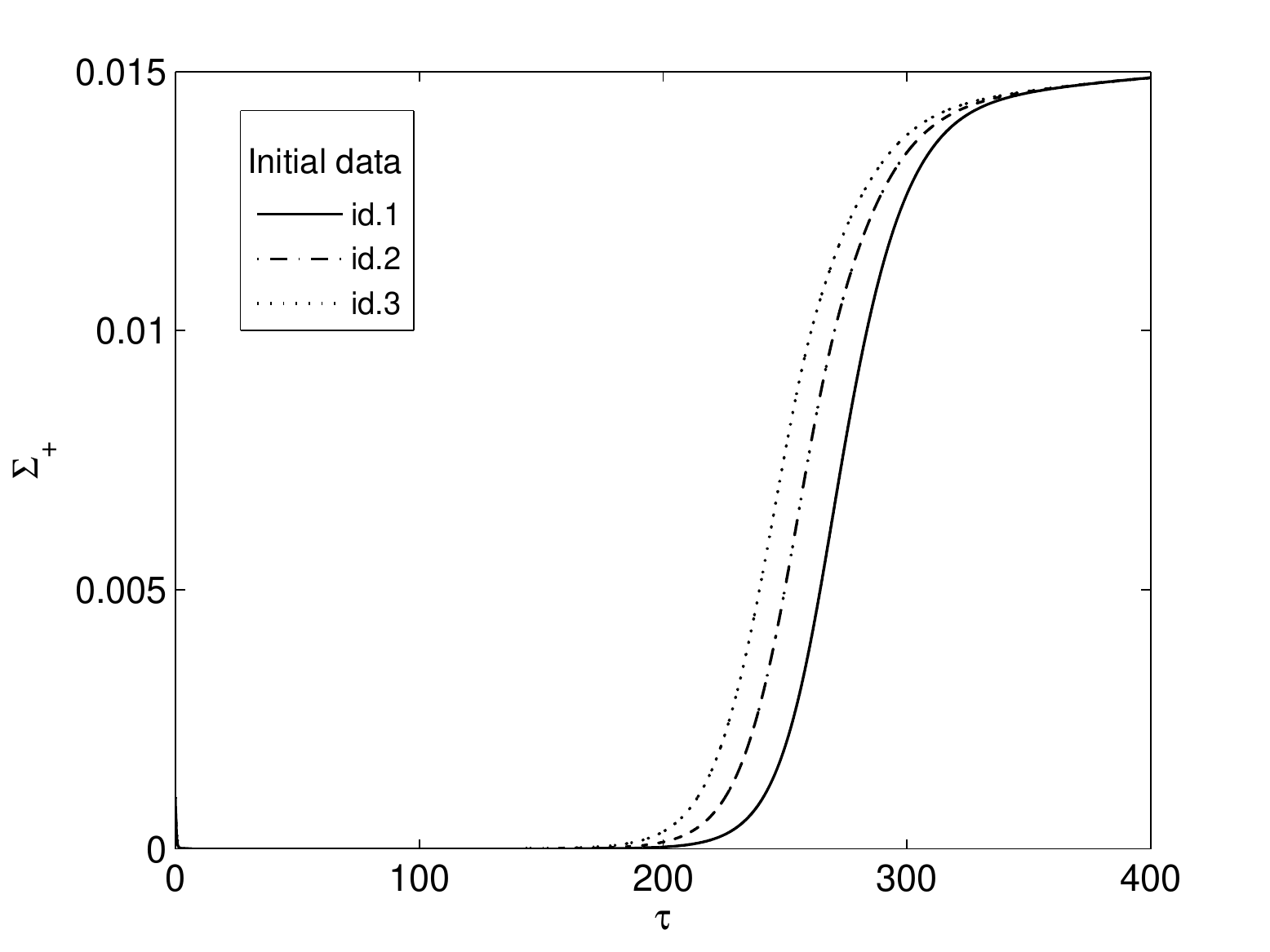}
    \caption{Anisotropy variable $\Sigma_+$ (BVIII). }
    \label{fig:BVIII:SigmaPlus}
  \end{minipage}
  \begin{minipage}{0.49\linewidth}
    \centering
    \includegraphics[width=\textwidth]{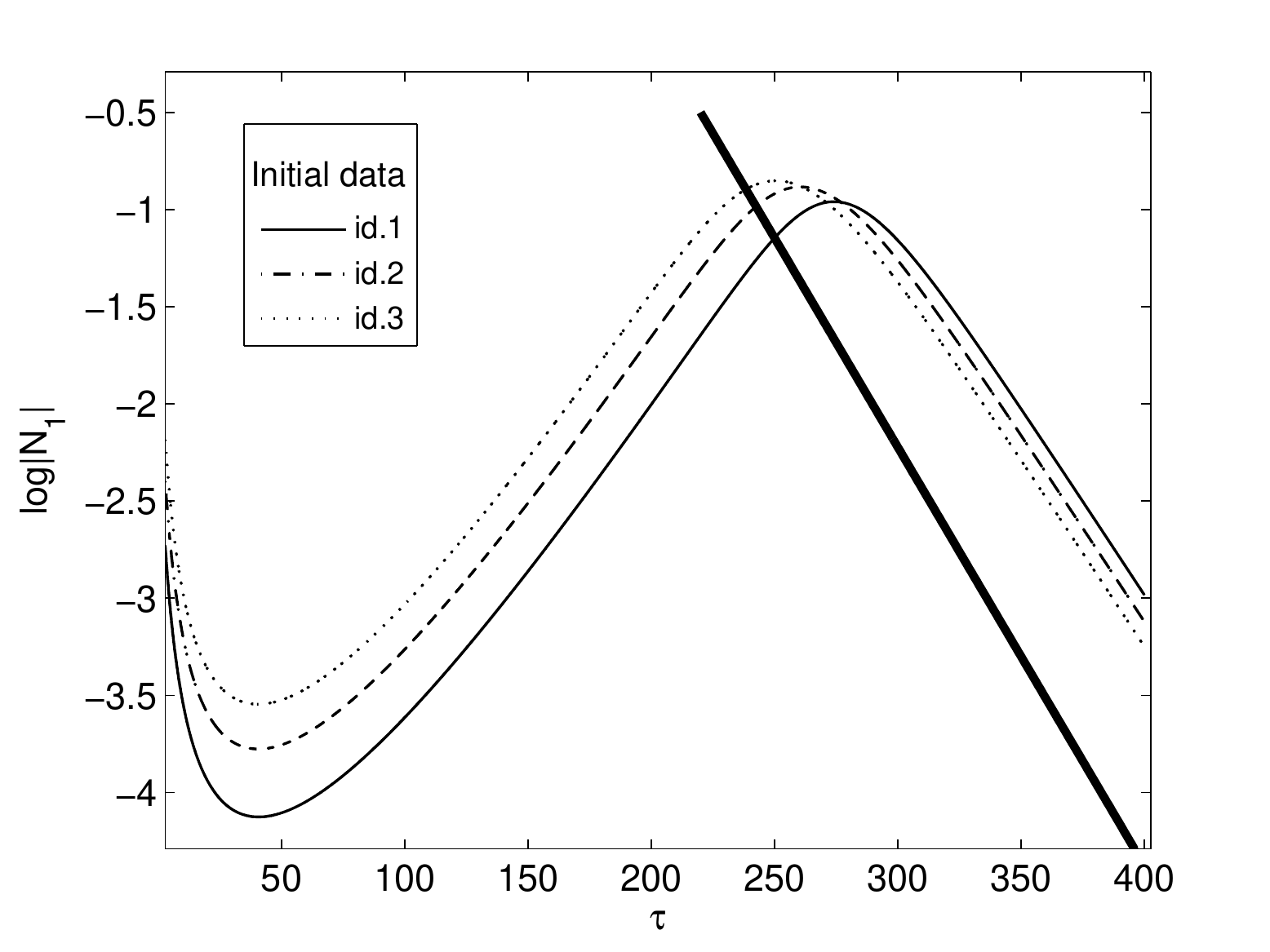}
    \caption{Decay of $N_1$ (BVIII). }
    \label{fig:BVIII:N1}
  \end{minipage}\hfill
  \begin{minipage}{0.49\linewidth}
    \centering
    \includegraphics[width=\textwidth]{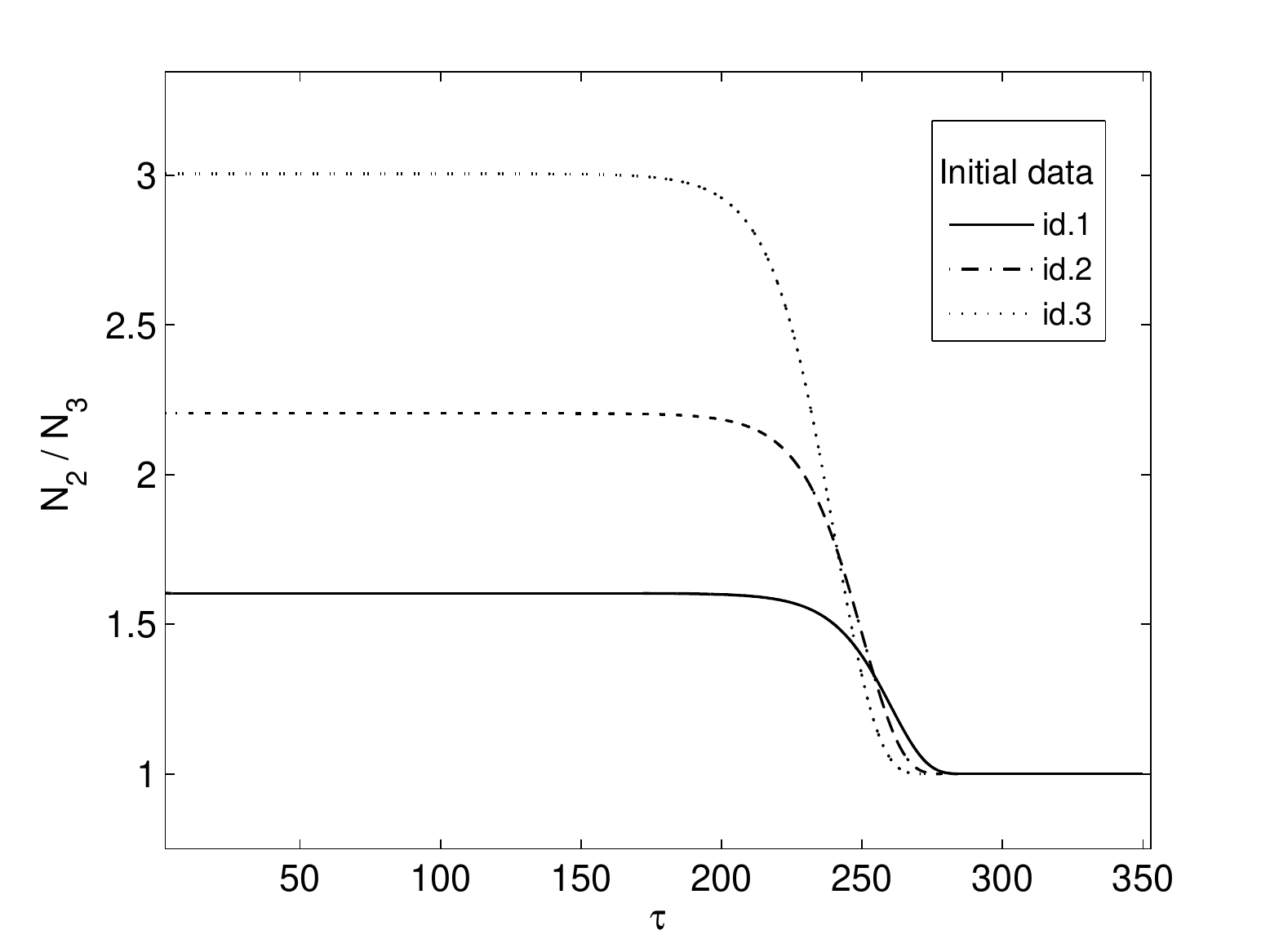}
    \caption{Evolution of $N_2/N_3$ (BVIII). }
    \label{fig:BVIII:N2N3}
  \end{minipage}
\end{figure}

\begin{figure}[t]
  \begin{minipage}{0.49\linewidth}
    \centering
    \includegraphics[width=\textwidth]{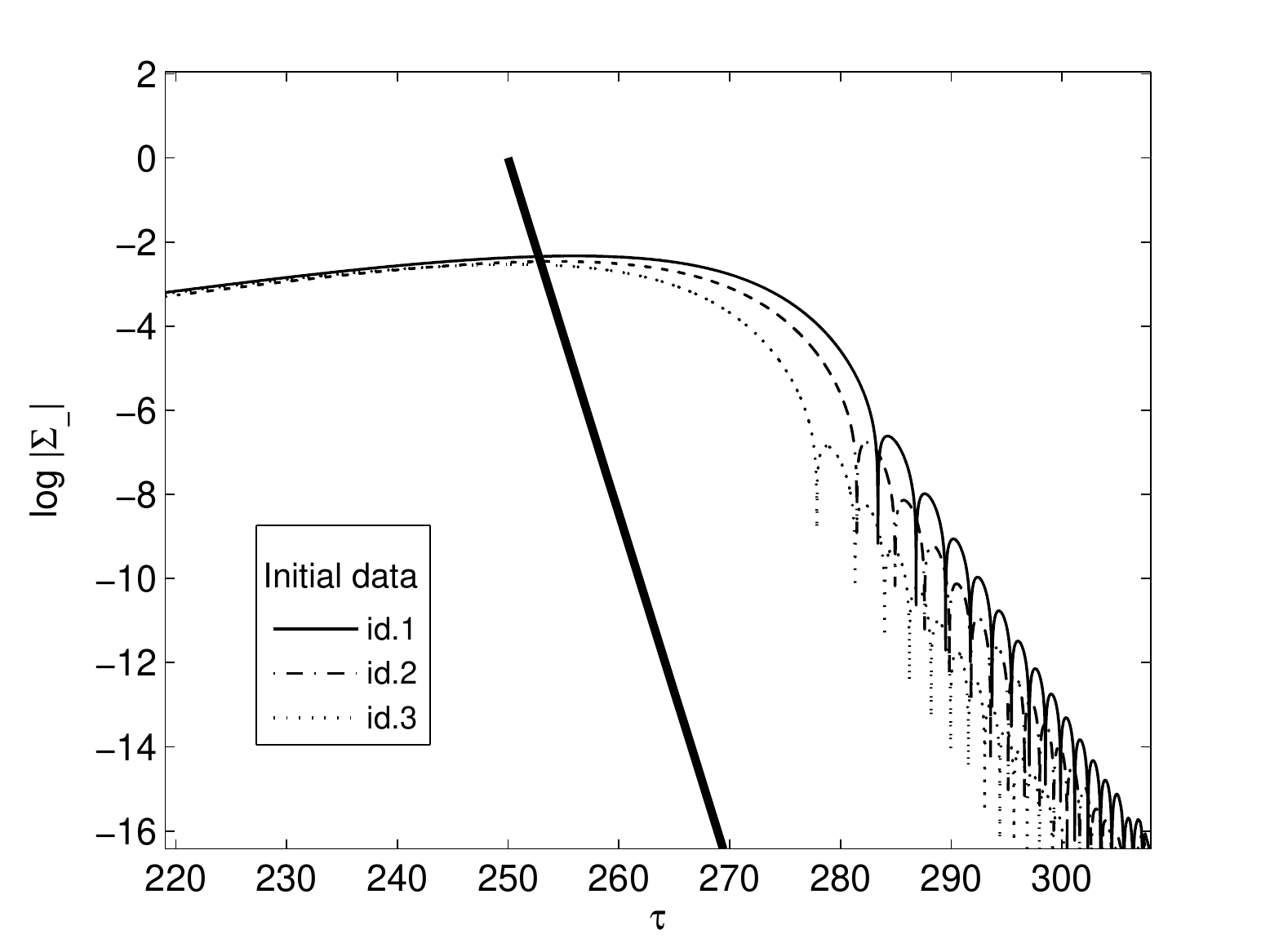}
    \caption{Decay of $\Sigma_-$ (BVIII). \vspace{2.8ex} }
    \label{fig:BVIII:SigmaMinus}
  \end{minipage}
  \begin{minipage}{0.49\linewidth}
    \centering  
    \includegraphics[width=\textwidth]{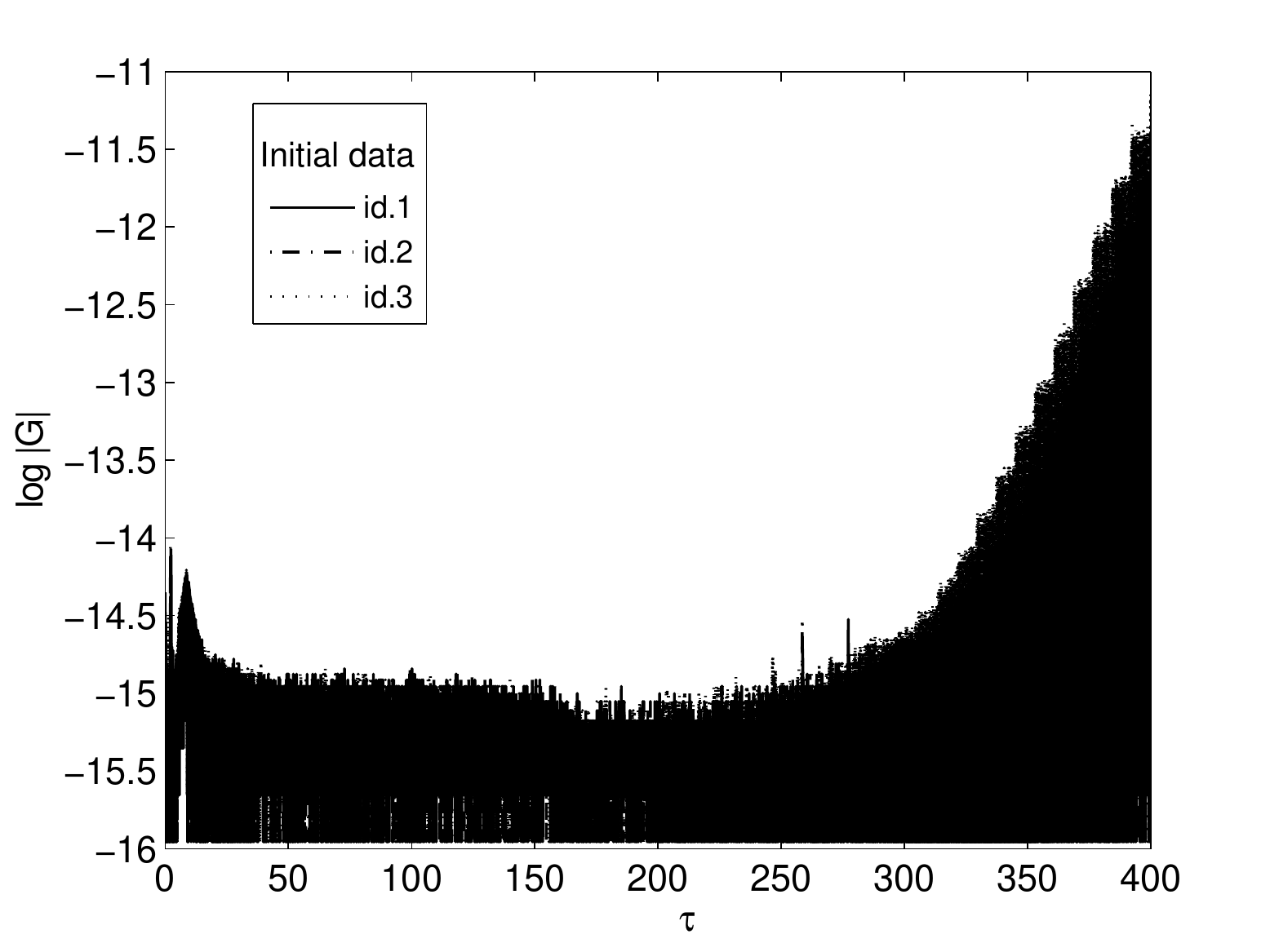}
    \caption{Constraint violations $G$ using the modified evolution equations (BVIII).}
    \label{fig:BVIII:G}
  \end{minipage}
\end{figure}

For our numerical studies, we choose $c_{1}= 1.45$ and $c_2=1.0$ and three different choices of initial data. This implies that $\Sigma_{+*}\approx 0.016$, $x_*\approx 0.57$, $y_*\approx 0.80$ and $q_*\approx 0.016$.
Indeed, the variables $\Sigma_\pm$, $x$, $y$, $N_1$ and $\psi$ approach these stationary values generically; see \Figsref{fig:BVIII:q} and \ref{fig:BVIII:SigmaPlus}. We remark that we would get an even better agreement with the theoretical predictions if we would have let the code run beyond $\tau=400$.
In any case, we find that generic initially inflationary Bianchi VIII solutions have a graceful exit from inflation.  In \Figref{fig:BVIII:N1}, we show the decay of $N_1$ which is in agreement with \Eqref{eq:BVIIIattractorN}; the bold line in \Figref{fig:BVIII:N1} represents the expected decay.
We demonstrate that $N_2-N_3\rightarrow 0$  in \Figref{fig:BVIII:N2N3}.
Similar to Bianchi VII$_0$, \Figref{fig:BVIII:SigmaMinus} shows that the decay of $\Sigma_-$ is oscillatory and that the decay rate is smaller than in the Bianchi I case (as represented by the bold line in the figure).  Just as a final check that our numerical code works reliably, we also plot the constraint violations in \Figref{fig:BVIII:G} for these runs. This plot demonstrates that our constraint damping scheme (see \Sectionref{sec:constraintdampingN}) works well for a long time. At some point, the numerical errors are dominated by the rapid growth of $N_2$ and $N_3$ and subtle cancellations which need to take place in order to compute $S_+$ and $S_-$, and hence the constraint violations start to grow. The time, however, for which the numerical solution is reliable, can be increased by decreasing the size of the numerical time step.

\paragraph{Results.}
The situation for Bianchi VIII is similar to Bianchi VII$_0$ as far as the behavior of the unbounded quantities $N_2$ and $N_3$ is concerned. Our numerical studies suggest that the other variables $\Sigma_\pm$, $x$, $y$, $N_1$ and $\psi$ are bounded and, in fact, approach stationary values asymptotically with certain universal values which only depend on $c_1$. The Bianchi VII$_0$ future attractor discussed in the previous section is not an attractor for Bianchi VIII solutions. Its behavior is quite different; for example, Bianchi VIII solutions do not isotropize and become spatially flat. In comparison of this and Theorem~1.2 in \cite{Ringstrom:2001es}, it turns out that the asymptotic properties of generic Bianchi VIII solutions are similar in the vacuum and the scalar field case
(in contrast to the Bianchi VII$_0$ case). Indeed, the same asymptotic behavior as in the vacuum case is obtained formally in the limit $c_1\rightarrow\infty$. 
In any case, our main claim that  generic initially inflationary  Bianchi VIII scalar field models have a graceful exit from inflation (if $c_1$ is chosen appropriately) has been confirmed.


\subsection{Beyond the monotonic case of the potential}
\label{sec:nonmonotonic}

\subsubsection{Monotonic potentials vs.\ potentials with a valley}
We recall the three main cases for the scalar field potential $V(\phi)$ given by \Eqref{eq:potential} in \Figref{fig:potential} which mainly differ in the number of critical points of the function $V$.  
 Notice that the critical points correspond to the zeros of the function $F(\phi)$ defined in \Eqref{eq:defFFirst} which are given by 
\begin{equation*}
\label{eq:criticalpoints}
\phi_{1,2}=\dfrac{1 \pm \sqrt{1-c^{2}_{1} c_2 } }{c_1} .
\end{equation*}
There are two (real) critical points --- a local minimum and a local maximum as for the third plot in \Figref{fig:potential} --- if and only if $1-c^{2}_{1} c_2> 0$. We shall refer to this case as the ``potential with a valley''. The potential has only one critical point --- a saddle point as in the second plot in the figure --- if and only if $1-c^{2}_{1} c_2=0$. We shall ignore this case in the following. Finally, there are no critical points --- as for the first plot in the figure --- if and only if $1-c^{2}_{1} c_2<0$; this is the ``monotonic'' case which we have considered exclusively so far.
 In this latter case, we have seen that the scalar field $\phi$ is able to roll down the potential hill all the way to infinity under generic conditions. This is why our asymptotic analysis based on the limit $\phi\rightarrow\infty$ (or equivalently $\psi=1/\phi\rightarrow 0$) discussed in the previous sections is relevant. On the other hand, if the potential has a valley, it is expected to depend on the initial data if the scalar field is able to go all the way to infinity. If it does, then we expect the same asymptotics as discussed for the monotonic potential. If it does not and is ``stuck'' in the valley instead, it is expected that the future asymptotics of the solutions is different. 

Let us consider the potential with a valley now. Since $F(\phi_1)=0$ at the minimum $\phi=\phi_1$ (as well as at the maximum), the condition $x'=0$ in \Eqref{eq:evolutioneqx} implies that $x=0$ (if we exclude the case $q=2$ for $y>0$ for the same reason as before). Then \Eqref{eq:evolutioneqy} together with $y'=0$ yields that $q=-1$. The remaining evolution equations hence yield the equilibrium point
\[\Sigma_\pm=0,\quad N_1, N_2, N_3=0,\quad x=0,\quad y=1,\quad \phi=\phi_1,\]
with $q=-1$.
The standard stability analysis reveals that this equilibrium point is hyperbolic and in particular future stable. Indeed, it is just the de-Sitter solution where the non-dynamic scalar field represents a positive cosmological constant.
The results in \cite{Rendall:2004gu} suggest that solutions for which the scalar field is trapped in the valley approach this equilibrium point asymptotically, and hence inflation never stops and is de-Sitter like asymptotically. This is indeed confirmed numerically below.  A similar analysis for the maximum $\phi=\phi_2$ of the potential yields that there exists no  stable equilibrium point --- in agreement with expectations.

\subsubsection{Numerical studies of the potential with a valley}

We choose the parameters $c_1=1.45$, $c_2=0.3$ which corresponds to the case of two critical points, i.e., the ``case with a valley''. The two critical points represent the locations of a local minimum at $\phi_1\approx 0.27$ and a local maximum at $\phi_2\approx 1.11$. For this discussion, we restrict to the simplest Bianchi case --- Bianchi I --- and explore numerically whether solutions starting from inflationary initial data have graceful exits or not. 

\begin{figure}[t]
  \begin{minipage}{0.49\linewidth}
    \centering 
    \includegraphics[width=\textwidth]{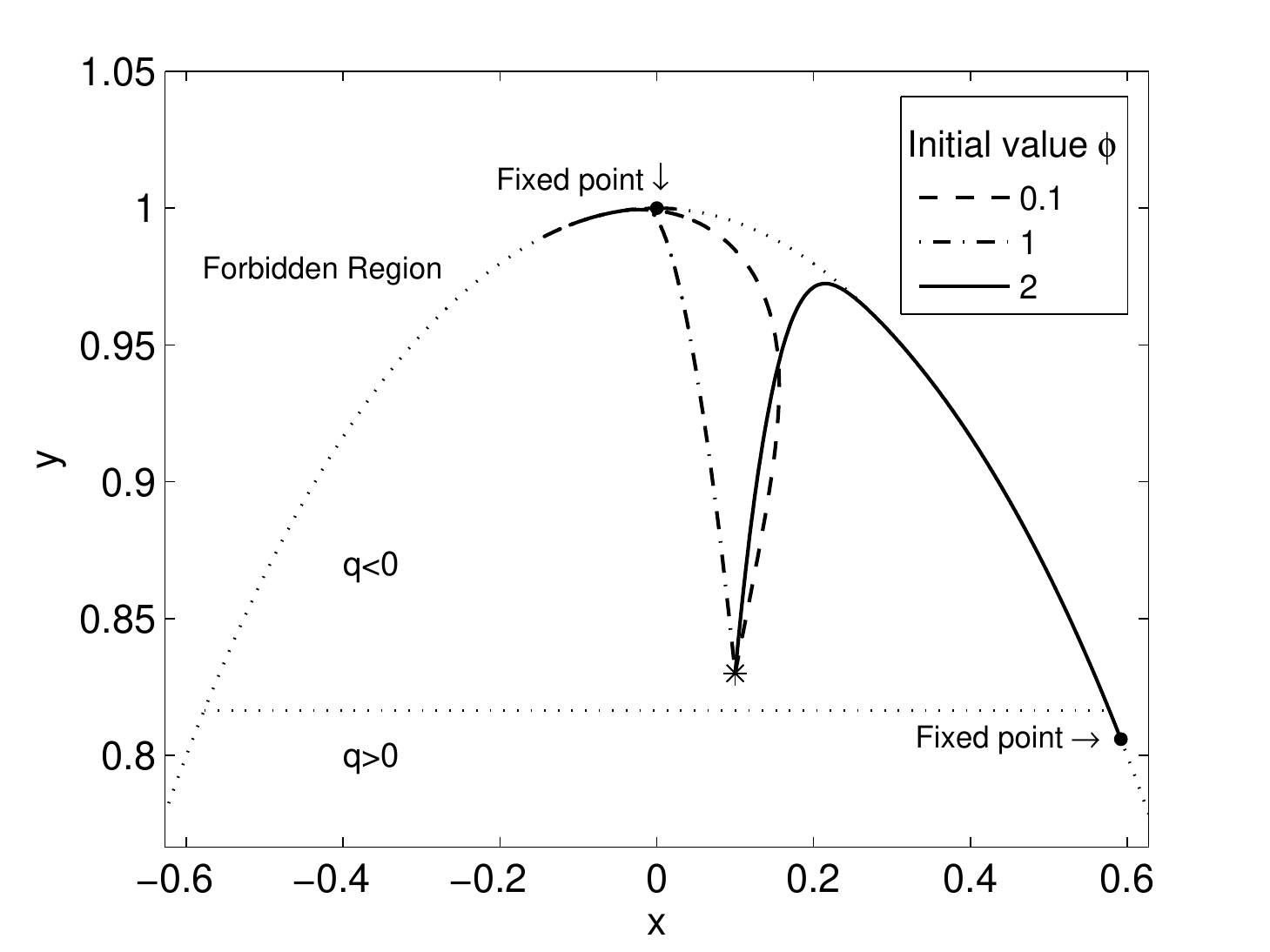}
    \caption{Valley potential: Projection of state space for different initial data of $\phi$ and fixed $x_0 = 0.1$ (BI).}
    \label{fig:BI:xyvarphi}
  \end{minipage}
  \hfill
  \begin{minipage}{0.49\linewidth}
    \centering 
    \includegraphics[width=\textwidth]{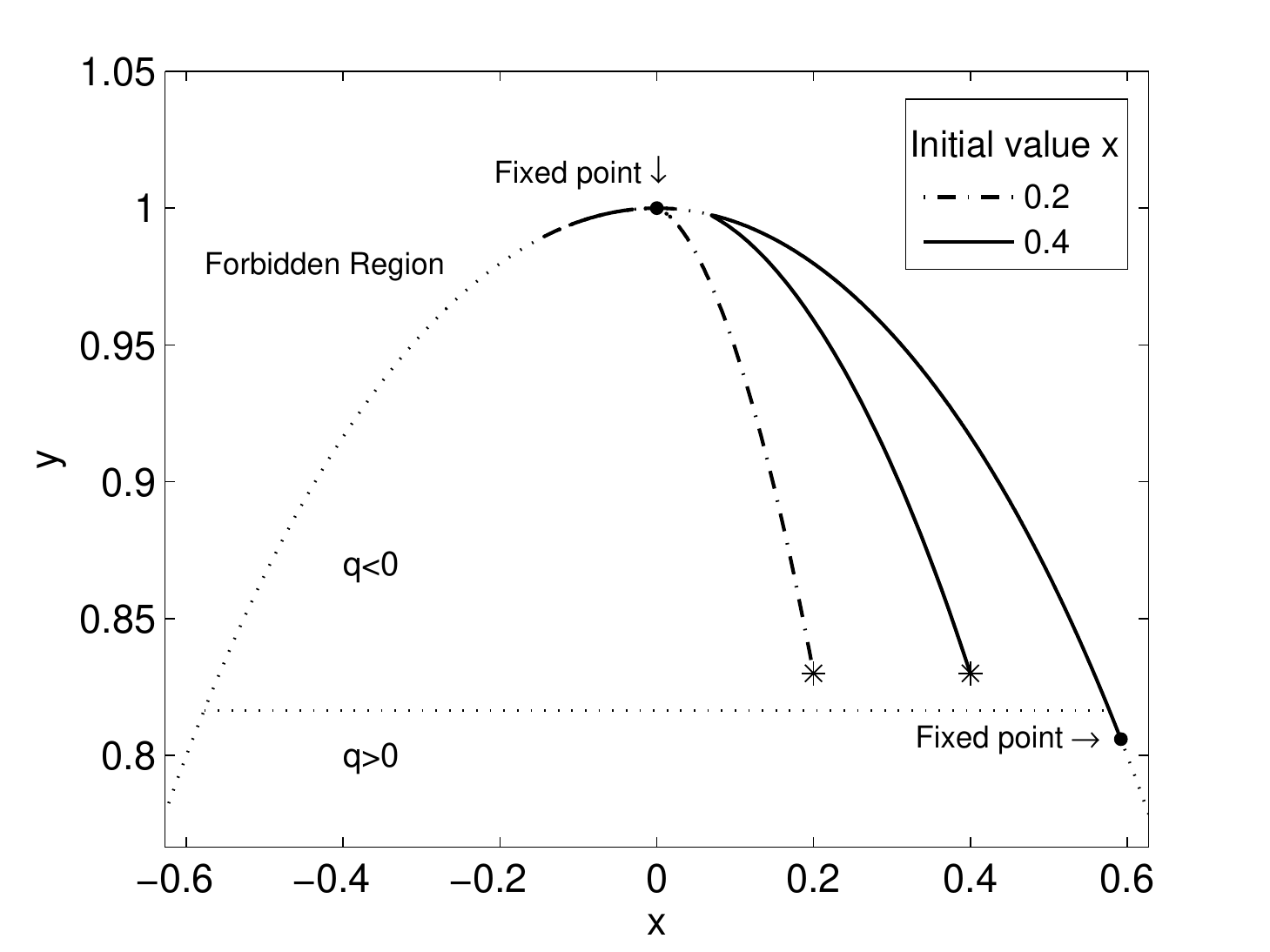}
    \caption{Valley potential: Projection of state space for different initial data of $x$ and fixed $\phi_0=1$ (BI).}
    \label{fig:BI:xyvarx}
  \end{minipage}
\end{figure}

In \Figref{fig:BI:xyvarphi}, we compute the projections of the orbits to the $x$-$y$-plane  in the state space (similar to \Figref{fig:BIxy}). We choose three different inflationary initial data, one where the initial value $\phi_0$ of the scalar field $\phi$ is $0.1$ and hence satisfies $\phi_0<\phi_1$ (on the ``downhill side'' inside the valley), one where $\phi_0$ has the value $1.0$ and hence satisfies $\phi_1<\phi_0<\phi_2$ (on the ``uphill side'' inside the valley) and one where $\phi_0$ has the value $2.0$ and hence satisfies $\phi_2<\phi_0$ (outside of the valley). All three choices of initial data have the same \textit{positive} initial value $x_0=0.1$ of $x$, and hence the scalar field increases initially (recall that $x$ is proportional to $\dot\phi$). The outcome is in agreement with our expectations, see \Figref{fig:BI:xyvarphi}. The solution corresponding to the first choice of initial data (notice that the solutions corresponding to all three initial data start from the same point in the $x$-$y$-plane of the state space) has the property that the scalar field roles down the downhill side of the valley and hence $x$ increases initially. It crosses the minimum of the potential with a positive (almost maximal velocity $x$)  and then roles up the uphill side inside the valley. This has the consequence that $x$ decreases and eventually becomes zero somewhere half way up this side of the valley. Then $x$ becomes negative and the scalar field starts to role down towards the minimum again. After a while the whole process repeats but with a smaller velocity. Because $q$ is negative during this whole process, our investigations in \Sectionref{sec:basicconsequencesN} show that the orbits must approach the circle $x^2+y^2=1$ in \Figref{fig:BI:xyvarphi} and hence isotropize. Hence during each period of oscillation around the minimum, the solution gets closer to the equilibrium point associated with the minimum above, i.e., the de-Sitter solution. This solution is therefore never able to escape from the inflationary regime and hence there is no graceful exit. A very similar dynamics is yielded for the second choice of initial data above. The only difference is that the scalar field roles up the uphill side of the valley at the beginning. However, because the initial velocity $x$ is not large enough, it is not able to escape from the valley and therefore has the same destiny as in the first case above: eventually the solution approaches the equilibrium point which represents the de-Sitter solution and there is therefore no graceful exit. In the third   case, the scalar field is already outside of the value initially and initially moves away from the valley towards infinity. There is hence nothing which could stop the scalar field from reaching infinity and hence from having the same asymptotics as discussed in the case of the monotonic potential in previous sections. This is confirmed by comparing the third curve in \Figref{fig:BI:xyvarphi} with the curves in \Figref{fig:BIxy}.

In \Figref{fig:BI:xyvarx} we study the case of three different sets of inflationary initial data, all with the same $\phi_0=1.0$ (i.e., on the uphill side inside of the valley). This time, however, we vary the initial value $x_0$ of $x$, i.e., the initial velocity of the scalar field. If the initial velocity is positive but small, the solution is trapped inside the valley as before and hence inflation never stops. If the initial velocity is positive and sufficiently large,  we expect that the solution is able to escape from the valley and hence reach infinity. This is confirmed in \Figref{fig:BI:xyvarx}. In particular, we notice that then the future asymptotics is the same as in the case of the strictly monotonic potential:  the solution approaches the same equilibrium point as above, and, in particular, there is a graceful exit from inflation.

\section{Discussion}
\label{sec:discussion}

In this paper we have investigated cosmological models with graceful exits
from inflation. We have restricted to minimally coupled self-gravitating Bianchi A scalar field models with a particular form of the potential. We have chosen the potential specifically in order to violate known sufficient conditions for eternal inflation in the literature. Most of the studies of the graceful exit problem in the literature are restricted to the spatially homogeneous and isotropic case; see for example \cite{Parsons:1995bb}, where the graceful exit problem has been considered for a very similar class of potentials  (in \Eqref{eq:potential}, one sets $c_2=0$ and the factor $\phi^2$ is replaced by $\phi^n$ for a general $n$). We find that the presence of future attractors in the decelerated regime, if the parameter $c_1$ of the potential is chosen specifically and if the scalar field ``escapes to infinity'' (which is always the case if our potential is monotonic), guarantees that initially inflating models become decelerated after a finite evolution time. In particular, they approach one of these attractors.  In the Bianchi cases I, II, VI$_0$, these future attractors are equilibrium points of the dynamical system for the Hubble normalized variables. In the Bianchi cases VII$_0$ and VIII, however, some of the geometric variables are unbounded and hence the asymptotic behavior is more difficult. Nevertheless, we are able to describe the asymptotic behavior of these solutions. Our investigations yield the following conclusions about the process of isotropization. While it is straightforward to prove that our models must isotropize and the spatial curvature must decay \textit{during} inflation, it depends on the Bianchi case whether this process continues or stops when inflation is over. In fact, for the Bianchi cases II and VI$_0$ and VIII, this process stops and anisotropies and spatial curvature grow again after inflation. One possible interpretation, given the remarkable level of isotropy of the cosmic microwave background, is that the Bianchi cases II and VI$_0$ and VIII do not constitute realistic models for our universe.
If the potential is not monotonic but has a ``valley'' instead then it depends on the particular choice of the initial conditions whether a graceful exit from inflation occurs. 

As we  have noticed before, we believe that some of the results in this paper generalize to more general classes of potentials; a particularly important example is the exponential potential. In fact, only little information about the potential is used in the analysis of the problem. This conjecture, however, needs to be studied in detail in future work. Also more work is required to understand the oscillatory behavior and the particular rates of decay in the Bianchi VII$_0$ and VIII cases.

Our techniques also allow us to determine the time between the begin of inflation and the graceful exit for our models. Physical arguments \cite{Mukhanov:2005ts} lead to the conclusion that inflation should end after $75-100$ efolds, which corresponds to the time $\tau=75-100$ for the graceful exit. The actual time when inflation stops for our models certainly depends on the particular choice of parameters and initial data. However, our investigations show that it is not necessary to ``fine-tune'' the parameters in order to achieve a graceful exit around this time. Another interesting related problem is the accelerated expansion of the \textit{present} universe due to \keyword{dark energy} which is suggested by current observations; see \cite{Komatsu:2011in} for recent observational data and \cite{Mukhanov:2005ts} for the theoretical background. The models, which we have presented in our paper, are not compatible with this since the decelerated epoch after the graceful exit lasts forever. However, as suggested by the results in \cite{Rendall:2004gu} and confirmed by further numerical investigations (which we do not present here), we find that if we add an arbitrarily small positive constant $V_*$ to our potential in \Eqref{eq:potential} (which is analogous to adding a positive cosmological constant to Einstein's equations), the corresponding solutions behave like the models presented in this paper for a long evolution time (in particular, inflation stops after a finite time). Then, however, at some late time which depends on the size of $V_*$, these models suddenly deviate from this and eventually approach the de-Sitter solution. Hence, by this slight modification,  we find the three main phases of the history of our universe according to the standard model: inflation, the intermediate decelerated epoch and the late time accelerated epoch of our present universe.

Our investigations are mainly based on numerical methods. Due to a constraint violating unstable mode near the solutions of interest and the particular form of the subsidiary system, it has been crucial for our investigations to make sure that the Friedmann constraint is satisfied as accurately as possible during the evolution. We have achieved this by introducing constraint damping terms to the evolution equations.

In this whole paper, we have excluded the Bianchi IX case. The main reason is that it is not guaranteed that the Hubble normalized variables, which we use, are well-defined. In fact, the Bianchi IX case should be studied independently, for example, using the variables introduced in \cite{Heinzle:2009bo}.

\section{Acknowledgments}

This work has been partly supported by the Marsden Fund Council from Government funding, administered by the Royal Society of New Zealand. The authors are grateful to A.~Rendall and H.~Friedrich for valuable discussions and insights. Moreover, we thank J.~Frauendiener for carefully reading the manuscript.

\bibliography{bibliography}

\end{document}